\documentclass[letterpaper]{JHEP3}
\usepackage{amsmath,amssymb}
\usepackage{graphicx}
\usepackage{comment}

\def\str{{\mathrm{str}}}

\title{\textbf{Universality of Mixed Action Extrapolation Formulae}}

\author{Jiunn-Wei Chen \\
Department of Physics and Center for Theoretical Sciences, National Taiwan University, Taipei 10617, Taiwan\\
E-mail: \email{jwc@phys.ntu.edu.tw}}

\author{Donal O'Connell\thanks{present address: School of Natural Sciences, Institute for Advanced Study, Princeton, NJ 08540} \\
California Institute of Technology, Pasadena, CA 91125, USA \\
E-mail: \email{donal@ias.edu}}

\author{ Andr\'e Walker-Loud \thanks{present address: Department of Physics, College of William and Mary, Williamsburg, VA 23187-8795, USA}\\
Department of Physics, University of Maryland, College Park, MD 20742-4111 \\
E-mail: \email{walkloud@wm.edu}}

\abstract{
Mixed action theories with chirally symmetric valence fermions exhibit very desirable features both at the level of the lattice calculations as well as in the construction and implementation of the low energy mixed action effective field theory.  In this work we show that when such a mixed action effective field theory is projected onto the valence sector, both the Lagrangian and the extrapolation formulae become universal in form through next to leading order, for all variants of discretization methods used for the sea fermions. Our conclusion relies on the chiral nature of the valence quarks. The result implies that for all sea quark methods which are in the same universality class as QCD, the numerical values of the physical coefficients in the various mixed action chiral Lagrangians will be the same up to lattice spacing dependent corrections.  This allows us to construct a prescription to determine the mixed action extrapolation formulae for a large class of hadronic correlation functions computed in partially quenched chiral perturbation theory at the one-loop level. For specific examples, we apply this prescription to the nucleon twist--2 matrix elements and the nucleon--nucleon system.  In addition, we determine the mixed action extrapolation formula for the neutron EDM as this provides a nice example of a $\bar \theta$-dependent observable; these observables are exceptions to our prescription.}

\begin{document}

\unitlength=1mm

\def\a{{\alpha}}
\def\b{{\beta}}
\def\d{{\delta}}
\def\D{{\Delta}}
\def\e{{\epsilon}}
\def\g{{\gamma}}
\def\G{{\Gamma}}
\def\k{{\kappa}}
\def\l{{\lambda}}
\def\L{{\Lambda}}
\def\m{{\mu}}
\def\n{{\nu}}
\def\w{{\omega}}
\def\O{{\Omega}}
\def\S{{\Sigma}}
\def\s{{\sigma}}
\def\t{{\tau}}
\def\th{{\theta}}
\def\x{{\xi}}

\def\ol#1{{\overline{#1}}}

\def\Dslash{D\hskip-0.65em /}
\def\dslash{{\partial\hskip-0.5em /}}
\def\vslash{{\rlap \slash v}}
\def\qbar{{\overline q}}

\def\CPT{{$\chi$PT}}
\def\QCPT{{Q$\chi$PT}}
\def\PQCPT{{PQ$\chi$PT}}
\def\tr{\text{tr}}
\def\str{\text{str}}
\def\diag{\text{diag}}
\def\order{{\mathcal O}}
\def\vit{{\it v}}
\def\vD{\vit\cdot D}
\def\am{\alpha_M}
\def\bm{\beta_M}
\def\gm{\gamma_M}
\def\smb{\sigma_M}
\def\smt{\overline{\sigma}_M}
\def\tb{{\tilde b}}

\def\mc#1{{\mathcal #1}}

\def\Bbar{\overline{B}}
\def\Tbar{\overline{T}}
\def\cBbar{\overline{\cal B}}
\def\cTbar{\overline{\cal T}}
\def\pq{(PQ)}

\def\eqref#1{{(\ref{#1})}}

%
%
\section{Introduction}

There has recently been a rapid growth in the use of \textit{mixed action} or \textit{hybrid} lattice QCD~\cite{Renner:2004ck,Edwards:2005kw} in the numerical computation of hadronic matrix elements~\cite{Bowler:2004hs,Bonnet:2004fr,Beane:2005rj,Edwards:2005ym,Beane:2006mx,Beane:2006pt,Beane:2006fk,Beane:2006kx,Alexandrou:2006mc,Beane:2006gj,Bar:2006zj,Hasenfratz:2006bq,Edwards:2006qx,Beane:2006gf,Orginos:2007tw,Hagler:2007xi}.  In response, there have been significant complementary developments in our theoretical understanding of mixed action (MA) lattice QCD through the use of mixed action effective field theory (EFT)~\cite{Bar:2002nr,Bar:2003mh,Tiburzi:2005vy,Bar:2005tu,Golterman:2005xa,Tiburzi:2005is,Chen:2005ab,O'Connell:2006sha,Bunton:2006va,Aubin:2006hg,Chen:2006wf,WalkerLoud:2006ub,Jiang:2007sn}.   Mixed action calculations allow one to use fermion discretization methods which respect chiral symmetry in the valence sector during the construction of the hadronic source and sink operators.  This is done in the background of numerically cheaper discretization methods in the sea sector (which generally explicitly violate chiral symmetry) during the generation of the gauge field configurations which contain the dynamical quark-antiquark polarization loops.  The main motivation stems from the significant numerical cost~\cite{Kennedy:2004ae,Edwards:2005an} of simulating either dynamical Kaplan (domain-wall) fermions~\cite{Kaplan:1992bt,Shamir:1993zy,Furman:1994ky} or dynamical overlap fermions~\cite{Narayanan:1992wx,Narayanan:1994gw} in the chiral regime as compared to Wilson fermions~\cite{Wilson:1974sk} (including clover~\cite{Sheikholeslami:1985ij} and twisted mass~\cite{Frezzotti:2000nk}) or staggered fermions~\cite{Kogut:1974ag,Susskind:1976jm}.  Kaplan and overlap fermions are often referred to as Ginsparg-Wilson (GW) fermions as they (approximately) satisfy the GW relation~\cite{Ginsparg:1981bj} and thus retain chiral symmetry on the lattice~\cite{Luscher:1998pqa} (modulo the quark masses).  The cost of these MA calculations employing GW valence fermions is then only the cost of propagator generation in the background of the dynamical gauge configurations (which presumably use one of the numerically cheaper varieties of fermions for the sea sector).  The most popular MA scheme~\cite{Bonnet:2004fr,Beane:2005rj,Edwards:2005ym,Beane:2006mx,Beane:2006pt,Beane:2006fk,Beane:2006kx,Alexandrou:2006mc,Beane:2006gj,Edwards:2006qx,Beane:2006gf,Orginos:2007tw,Hagler:2007xi} was developed by the LHP Collaboration~\cite{Renner:2004ck,Edwards:2005kw} in which domain-wall valence propagators are generated on the asqtad-improved~\cite{Orginos:1998ue,Orginos:1999cr} publicly available MILC configurations~\cite{Bernard:2001av}. 

In a recent paper~\cite{Chen:2006wf}, we showed that the chiral symmetry of the valence fermions suppresses sources of chiral symmetry breaking arising from the sea sector such that for many mesonic observables, there are no lattice spacing dependent counterterms through next-to-leading order (NLO) in the MA EFT.  Furthermore, this valence chiral symmetry is strong enough to suppress all explicit lattice spacing dependence of these mesonic observables, with the exception of modifications to the correlation functions arising from the unphysical hairpin contributions. This exception is due to the lack of unitarity inherent in MA or partially quenched (PQ) calculations~\cite{Bernard:1993sv,Sharpe:1997by}. These properties are only transparent when one uses an on-shell renormalization scheme, expressing correlation functions in terms of lattice-physical parameters measured directly in the calculation, such as the pion mass or decay constant, $m_\pi$ or $f_\pi$; in much of our discussion below we will assume that this renormalization scheme has been utilized. 

In this paper, we build on these results and extend them such that we can formulate a convenient prescription for converting quantities computed in partially quenched theories to expressions valid in mixed action theories. Our prescription is valid for a wide range of observables in a range of mixed action theories. To be clear, we state our requirements here and will discuss them more extensively throughout this work. In terms of the mixed action theory, we require that the hairpin structure in the mixed action theory is the same as in the partially quenched theory, and we require that the valence quarks are chiral. In terms of the observables, we require that there is no dependence on the CP violating $\bar \theta$ parameter. We will discuss these requirements below and later in Sec.~\ref{sec:EDM} we will describe how the neutron EDM, which of course depends on $\bar \theta$ requires a modification of our prescription.

We begin by observing that mixed action EFTs describing the light mesons have one unphysical operator appearing at leading order (LO) which is universal in form, regardless of the discretization used in the valence or sea sector, with only the numerical value of the coefficient depending on the actions used; the coefficient is known as $C_\mathrm{Mix}$ in the literature.%
\footnote{This parameter, or equivalently the mixed meson mass renormalization has recently been calculated for domain-wall valence and the coarse MILC staggered fermions~\cite{Orginos:2007tw}.} 
In Section~\ref{sec:MAEFT} we prove that for vertices with $2N$ mesons from this operator, for which two of the mesons are of a mixed valence-sea type, and the rest are purely valence (or sea), this operator functions identically to the LO operator involving the quark mass term.  Also in Sec.~\ref{sec:MAEFT}, we construct the most general MA Lagrangian projected onto the valence sector of the theory. This is particularly relevant because for correlation functions computed with valence fermions, we only need the counterterms of the valence sector. 

Our results imply that a large cancellation of potential lattice spacing dependent counterterms occurs in MA theories. This means that MA extrapolation formulae have a continuum functional form with only slight modifications, when expressed in lattice-physical parameters. Combined with our work in Section~\ref{sec:MAEFT}, this allows us to construct the general prescription we alluded to above. The prescription converts PQ chiral extrapolation formulae through the one-loop level into the corresponding MA extrapolation formulae, allowing one to make use of the extensive literature on partial quenching. As we will discuss in some detail in Sec.~\ref{sec:MAEFT}, our prescription requires three key components to be valid; the valence fermions are (approximately) chirally symmetric modulo the quark masses, the hairpin structure of the theory is the same as in partially quenched chiral perturbation theory (PQ$\chi$PT)~\cite{Bernard:1993sv,Sharpe:1997by,Sharpe:2000bc,Sharpe:2001fh} and the $\bar \theta$ term is negligible.

In Section~\ref{sec:appl}, we explicitly determine the MA formulae of several observables which are non-trivial examples of our prescription and are of current interest; Sec.~\ref{sec:EDM}, the neutron EDM which provides a nice example and requires a slight modification of our prescription, in Sec.~\ref{sec:gA} nucleon twist-2 matrix elements and in Sec.~\ref{sec:NN}, nucleon--nucleon scattering. In Section~\ref{sec:concl} we comment on our results and conclude. In the appendix, we describe why mixed action theories involving a twisted mass sea satisfy the requirements of our prescription.

%
%
\section{MA Effective Field Theory \label{sec:MAEFT}}

Mixed action EFT is a natural generalization of PQ$\chi$PT~\cite{Bernard:1993sv,Sharpe:1997by,Sharpe:2000bc,Sharpe:2001fh} reducing to it in the continuum limit.%
\footnote{For an introduction to MA EFT see Refs.~\cite{Bar:2002nr,Bar:2003mh,Tiburzi:2005vy,Bar:2005tu,Golterman:2005xa,Tiburzi:2005is,Chen:2006wf}.} 
Partially quenched $\chi$PT is constructed from the underlying theory, partially quenched QCD (PQQCD), analogously to how $\chi$PT~\cite{Weinberg:1978kz,Gasser:1983yg,Gasser:1984gg} is constructed from QCD.  Partially quenched QCD exhibits an approximate graded chiral symmetry (for light quark masses, $m_Q \ll \L_{QCD}$), with $N_v$ valence and $N_v$ ghost quarks and $N_s$ sea quarks,%
\footnote{See Ref.~\cite{Sharpe:2001fh} for a complete discussion of the PQ symmetry group.} 
\begin{equation*}
	SU(N_v+N_s|N_v)_L \otimes SU(N_v+N_s|N_v)_R \otimes U(1)_V\, , 
\end{equation*}
which is explicitly broken by the quark mass terms, $m_Q$.  It is then assumed, as with QCD, that the vacuum of PQQCD spontaneously breaks this symmetry down to the vector subgroup, giving rise to the PQ pseudo-Goldstone modes.  The PQ$\chi$PT Lagrangian is then constructed with a spurion analysis such that all operators respect the symmetries of  PQQCD.  These symmetries are further broken explicitly by mixed action effects.  At finite lattice spacing, $a$, there is no symmetry which mixes the valence and sea quarks, breaking the symmetry to a direct product of valence and sea sectors,
\begin{multline}
	SU(N_v+N_s|N_v)_{L} \otimes  SU(N_v+N_s|N_v)_{R}
	\\
	\, \underbrace{\longrightarrow}_{a\neq 0}\, 
	SU(N_v|N_v)_{L} \otimes SU(N_v|N_v)_{R} \otimes SU(N_s)_{L} \otimes SU(N_s)_{R}\, .
\label{eq:symmbreaking}
\end{multline}
For sufficiently small lattice spacing compared to the non-perturbative scale, these effects are perturbative and can be treated in an EFT framework.  All operators in the MA Lagrangian which do not explicitly depend upon the lattice spacing are given by their PQ equivalents, with the same value of the corresponding low energy constants (LECs).  There will be new operators with explicit lattice spacing dependence, some of which arise from the mixed action effects and break the PQ symmetry, and whose exact form depends upon the lattice actions used.  These new \textit{unphysical} operators will contribute to correlation functions of observable quantities, for which the extrapolation formulae can be determined from the appropriate mixed action EFT and then used to remove these unphysical contaminations from MA lattice QCD calculations.

However, mixed action theories that have chirally symmetric valence fermions, such as domain-wall fermions in the infinite $5^{th}$ dimension limit, or overlap fermions with a perfect overlap operator, give rise to chiral extrapolation formulae for a large class of valence observables which are \textit{identical} in form through next-to-leading order, with any and all discretization methods used for the sea fermions.  Provided the various sea quark discretization methods are in the same universality class as QCD, and that the lattice spacing dependent chiral symmetry breaking is perturbative\footnote{Notice that if the lattice spacing dependent sources of chiral symmetry violation are too large, then chiral perturbation theory is simply not the correct effective field theory describing the low energy dynamics of the underlying lattice theory. Similarly, chiral perturbation theory would not be relevant in nature if quark masses were all large compared to the scale of chiral symmetry violation.}, that is, the lattice spacing $a$ is small compared to the scale of chiral symmetry breaking $\Lambda_\chi$ in the sense that $a \Lambda_\chi \ll 1$, in the same way that the quark masses $m_q$ give rise to perturbative chiral symmetry breaking since $m_q \ll \Lambda_\chi$, the only difference between these various mixed action theories will be in the numerical value of the \textit{unphysical} counterterms which enter the chiral extrapolation formulae.  Furthermore, these extrapolation formulae are sufficiently continuum like, due to the good chiral properties inherited by the chiral symmetry of the valence fermions, that they can be determined from the corresponding formulae in partially quenched $\chi$PT.  In the rest of Sec.~\ref{sec:MAEFT}, we present the formalism necessary to understand this claim and then provide our mixed action prescription.

%
%
\subsection{Matching the $\mc{O}(a^{2})$ operators \label{sec:Spurions}}

To construct the mixed action effective Lagrangian, one must first construct the continuum Symanzik quark level Lagrangian~\cite{Symanzik:1983dc,Symanzik:1983gh} which respects all the symmetries of the underlying lattice action.  Then one performs a spurion analysis on this continuum Lagrangian to determine the operators in the mixed action EFT~\cite{Sharpe:1998xm}.%
\footnote{We note that it is unnecessary to consider the color structure of these effective continuum quark level operators as this does not play a role in the pattern of symmetry breaking used to construct the EFT operators.} 
A specified power counting orders the infinite tower of operators entering the low energy Lagrangian.  In this work we consider the general small parameter to be
\begin{equation}\label{eq:power_counting}
	\varepsilon^2 \sim \frac{m_\pi^2}{\L_\chi^2} \sim a^2 \L_\chi^2\, .
\end{equation}
The order at which the LO lattice spacing dependent operators appear depends upon the specific action used.  However, to determine the counterterm structure of the chiral Lagrangian relevant to valence quantities through NLO in the mixed action EFT, we only need to understand the quark level operators of the valence and mixed sectors of the theory.  The sea quark operators will be important for determining the additive mass renormalization of the sea-sea mesons, which is important for understanding the hairpin interactions, but otherwise the sea quark Symanzik operators will only lead to trivial renormalizations of counterterms relevant to valence quantities.  This will become more clear in our discussion below.  We also stress that this only holds through the leading loop order in the EFT, after which the extrapolation formulae for valence quantities will become dependent upon the details of the underlying lattice action in the sea sector.

The chiral symmetry of the valence sector prohibits operators of dimension five in the Symanzik action~\cite{Bar:2002nr,Bar:2003mh,Bar:2005tu}.  Therefore, we begin with the dimension-6 operators which mix the valence and sea fermions, and are therefore necessarily four quark operators.  There will be $\mc{O}(a^2)$ operators in the valence sector but they will not break chiral symmetry because of the good chiral properties of the valence fermions.  These will then be indistinguishable from the operators already in the chiral Lagrangian and will amount to renormalizations of the physical operators.  There are also operators which break Lorentz symmetry, but are singlets under the hypercubic group which we do not consider here as they are generally higher order than we are working to.  In Ref.~\cite{Tiburzi:2005vy}, these Lorentz violating operators have been analyzed for the baryons.  As we mentioned above there will also be $\mc{O}(a^2)$ operators in the sea-sector,%
\footnote{For unimproved Wilson sea fermions, there will be $\mc{O}(a)$ operators.  This leads to an extra complication with the hairpin interactions, but otherwise does not modify our prescription.  We will address this in more detail in Sec.~\ref{eq:MALagrangian}.} 
but these transform as singlets under chiral rotations of the valence fermions, and thus can be absorbed into generic $\mc{O}(a^2)$ operators needed to renormalize purely valence correlation functions.%
\footnote{Of course there are exceptions to this rule, but these occur only in special cases which are related to the class of quantities which do not follow our discussion and prescription.  We will postpone the discussion of these special cases to Sec.~\ref{sec:EDM}, where we will discuss them in the context of the neutron EDM.} 
Therefore, to construct the mixed action chiral Lagrangian we only need to consider the mixed valence-sea operators at this order.  Much of this discussion can be found in Refs.~\cite{Bar:2002nr,Bar:2003mh,Tiburzi:2005vy,Bar:2005tu,Tiburzi:2005is,Chen:2006wf}.

The valence-sea mixing Lagrangian at dimension six for chirally symmetric valence fermions and any type of sea fermion is given by four-quark operators which are products of valence and sea quark-bilinears that independently respect chiral symmetry in the valence and sea sectors respectively~\cite{Bar:2002nr,Bar:2003mh,Tiburzi:2005vy,Bar:2005tu,Tiburzi:2005is}.  However, because these operators explicitly break the partially quenched symmetry relating the sea and valence fermions, despite the fact that they are constructed from chirally symmetric quark-bilinears, they give rise to additive mass corrections for mixed hadrons composed of both valence and sea fermions.  This, for example, is how a mixed pion of domain-wall valence and sea quarks, but with different values of the Wilson-``r" parameter, or different $5^{th}$ dimensional extent, are subject to additive lattice spacing dependent mass corrections.  In the non-mixed action limit, these operators are no longer allowed by the symmetries of the lattice action and must vanish.  At dimension-6, there are only two allowed mixed action operators which are both 4-quark operators,
\begin{align}\label{eq:L6}
	{\mc{L}}_{\mathrm{Mix}}^{(6)} =&\ a^2 C_{\mathrm{Mix}}^{V}\, \left( 
		\overline{Q}\, \g _{\mu } \mc{P}_{V}\, Q \right) 
		\left( \overline{Q}\, \g ^{\mu } \mc{P}_{S}\, Q \right) 
	+a^{2} C_{\mathrm{Mix}}^{A} \left( 
		\overline{Q}\, \g_{\mu }\g _{5}\mc{P}_{V}\, Q \right) 
		\left( \overline{Q} \g ^{\mu}\g _{5}\mc{P}_{S}\, Q \right)
	\nonumber\\
	=&\ 
	a^2 C_{\mathrm{Mix}}^{LL} \left( \overline{Q}_L\, \g _{\mu }\mc{P}_{V}^{L}\, Q_L \right)
		\left( \overline{Q}_{L}\, \g _{\mu }\mc{P}_{S}^{L}\, Q_{L} \right) 
	+a^2 C_\mathrm{Mix}^{LR} \left( \overline{Q}_L\, \g _{\mu }\mc{P}_{V}^{L}\, Q_{L} \right)
		\left( \overline{Q}_{R}\, \g _{\mu }\mc{P}_{S}^{R}\, Q_{R} \right)
	\nonumber\\
	&\ +\left[ L\longleftrightarrow R \right]\, ,
\end{align}
where $\mc{P}_V$ and $\mc{P}_S$ are valence and sea projection operators respectively.  The coefficients $C_{\mathrm{Mix}}^{LL}=C_{\mathrm{Mix}}^{RR}$ and $C_{\mathrm{Mix}}^{LR}=C_{\mathrm{Mix}}^{RL}$ due to parity conservation.  The introduction of $\mc{P}_{V(S)}^{L}$ and $\mc{P}_{V(S)}^{R}$ is convenient for spurion analysis after which one can set 
$\mc{P}_{V(S)}^{L}=\mc{P}_{V(S)}^{R}=\mc{P}_{V(S)}$.  Under chiral transformations, 
$Q_{L}\rightarrow LQ_{L}$ and $Q_{R}\rightarrow RQ_{R}$.  Equation~\eqref{eq:L6} will be invariant under these transformations if 
\begin{equation}\label{eq:P_VS}
	\mc{P}_{V(S)}^{L} \rightarrow L\, \mc{P}_{V(S)}^{L}\, L^{\dagger },\ 
	\mc{P}_{V(S)}^{R} \rightarrow R\, \mc{P}_{V(S)}^{R}\, R^{\dagger }.
\end{equation}%
In MA$\chi$PT, the hadronic fields transform as in PQ$\chi$PT under the chiral transformations, which are~\cite{Labrenz:1996jy,Savage:2001jw,Chen:2001yi,Beane:2002vq},
\begin{align}
	\Sigma(x) &\rightarrow L\, \Sigma(x)\,  R^{\dagger}\quad,\quad 
	\xi(x) \rightarrow L\, \xi(x)\,  U^{\dagger}(x) = U(x)\, \xi(x)\, R^{\dagger}
	\nonumber\\
	\mc{B}_{ijk} &\rightarrow 
		(-1)^{\eta_l(\eta_j+\eta_m) +(\eta_l+\eta_m)(\eta_k+\eta_n)}
		U(x)_i^{\ l} U(x)_j^{\ m} U(x)_k^{\ n} \mc{B}_{lmn}\, ,
\end{align}
where $\S$ contains the meson fields, $\xi = \sqrt{\S}$, $\mc{B}$ is a spin-$1/2$ baryon field and $U(x)$ is a complicated transformation which depends upon the mesons, and thus on spacetime.  The chiral transformations of the spin-$3/2$ fields, $\mc{T}$ are identical to those of $\mc{B}$ and the heavy meson transformations can be found in Ref.~\cite{Savage:2001jw}.  The grading factors, $\eta_i$ keep track of the (anti)commuting nature of the different quark fields, where the sea and valence quarks are anti-commuting and the ghost quarks are commuting.  For our discussion we will be mostly interested in products of fields with purely valence fermions as these will be the relevant degrees of freedom to construct the counterterms for valence quantities at NLO.  With this restriction, it is often more convenient to think about how the valence fields transform.  The valence--nucleon field $N_V$, for example, can be related to the valence projected $\mc{B}$ field and transforms more simply under the reduced chiral transformations
\begin{align}
	\mc{B}_{ijk} &= \frac{1}{\sqrt{6}} \left( \varepsilon_{ij}N_k + \varepsilon_{ik}N_j \right)
		\qquad \textrm{ for $i,j,k \in$ valence} \\
	N_V &\rightarrow U(x) N_V\, .
\end{align}
The projection operators transform as in Eq.\eqref{eq:P_VS} and thus we also have
\begin{align}
	\left( \xi ^{\dagger }\mc{P}_{V(S)}^{L}\xi \right)  &\longrightarrow
		U\left( \xi ^{\dagger }\mc{P}_{V(S)}^{L}\xi \right) U^{\dagger}, 
	\notag \\
	\left( \xi \mc{P}_{V(S)}^{R}\xi ^{\dagger }\right)  &\longrightarrow
		U\left( \xi \mc{P}_{V(S)}^{R}\xi ^{\dagger }\right) U^{\dagger}\, ,
\end{align}
which allows us to define the projectors 
\begin{equation}
	\mc{P}_{V(S)}^{\pm } = \frac{1}{2} \left( \xi ^{\dagger }\mc{P}_{V(S)}^{L}\xi 
		\pm \xi \mc{P}_{V(S)}^{R}\xi ^{\dagger} \right)\, ,
\end{equation}
with 
\begin{equation}
\mc{P}_{V(S)}^{\pm }\rightarrow U\mc{P}_{V(S)}^{\pm }U^{\dagger }\, ,
\end{equation}
which are even (+) and odd (-) under parity.  Now we can construct the set of $\mc{O}(a^2)$ operators in the MA chiral Lagrangian relevant for valence quantities, which must have the same chiral symmetry properties as Eq.~\eqref{eq:L6},
\begin{equation}
	\mc{L}_{a^{2}}^{\mathrm{Mix}} =a^{2}\left( \mc{V}_{\phi}^\mathrm{Mix} 
		+\mc{V}_{N}^\mathrm{Mix} 
		+\mc{V}_{NN}^\mathrm{Mix}
		\right)\, .
\end{equation}
The mixed action meson potential was first determined in Ref.~\cite{Bar:2003mh}. We will deduce the potential here on the basis of symmetry considerations in order to emphasize its universality. The symmetry breaking of Eq.~\eqref{eq:symmbreaking} enlarges the number of operators which we can form from the meson field $\Sigma$, and which are invariant under the symmetry. These operators occur at order $\mc{O}(a^2)$, which we take to be of order $m_q$ in our power counting. Consequently, the lowest order operators which are invariant under the reduced symmetry are
\begin{align}
	\mathcal{O}_1 &= a^2 \mathrm{str} (\mc{P_S} \Sigma \mc{P_S} \Sigma^\dagger) ,
	\quad
	\mathcal{O}_2 = a^2 \mathrm{str} (\mc{P_V} \Sigma \mc{P_S} \Sigma^\dagger) ,
	\\
	\mathcal{O}_3 &= a^2 \mathrm{str} (\mc{P_S} \Sigma \mc{P_V} \Sigma^\dagger) ,
	\quad
	\mathcal{O}_4 = a^2 \mathrm{str} (\mc{P_V} \Sigma \mc{P_V} \Sigma^\dagger) .
\end{align}
However, using the identities $\mc{P_S} + \mc{P_V} = \mathbf{1}$ and $\Sigma^\dagger \Sigma = \mathbf{1}$, it
is easy to see that these operators are equal up to constant numeric factors, so that there is only one non-trivial operator invariant under the reduced symmetry group at this order. Thus, the mixed action meson potential is conventionally given by
\begin{equation}\label{eq:VMix}
	\mc{V}_{\phi}^\mathrm{Mix} 
	= C_\mathrm{Mix}\, \str \left( T_3 \Sigma T_3 \Sigma^{\dagger} \right)
\end{equation}%
with
\begin{equation}\label{eq:T3}
	T_3 = \mc{P}_S - \mc{P}_V\, .
\end{equation}
Of course, the quark level Lagrangian given in Eq.~\eqref{eq:L6} is universal by the same reasoning.  Note, this also applies to unimproved Wilson sea fermions as well.

There are additional $\mc{O}(a^{2})$ mixed contributions involving the sea
quarks.  These operators are singlets under chiral rotations of the
valence quarks so they can be absorbed into the PQ coefficients and, in
general, will give additive $a^{2}$ contributions to the PQ coefficients.  For example,  
\begin{align}\label{eq:NMixCounterterm}
	\mc{V}_{N}^\mathrm{Mix} &= C_{N}^{\prime } \left( \overline{N}_{V}\mc{P}_{V}^{+}N_{V} \right) 
		\mathrm{str}[\mc{P}_{S}^{+}]
	\notag \\
	&\longrightarrow C_{N} \overline{N}_{V}N_{V}\, ,
\end{align}
which we see is an additive correction the valence nucleon mass.%
\footnote{The operator $\left( \overline{N}_{V}\mc{P}_{S}^{+}N_{V}\right) \mathrm{str}[\mc{P}_{V}^{+}]$ does not contribute because $\mathrm{str}[\mc{P}_{V}^{+}]=0$.} 
Therefore, we can just focus on the $\mc{O}(a^{2})$ contributions of the valence fields from now on as the contributions from the mixed terms are indistinguishable under valence-chiral rotations.  If we are interested in matrix elements with no pions in the external states, the nucleon counterterms behave as if $\xi =\xi^{\dagger }=1$ and therefore
\begin{equation}\label{proj}
	\mc{P}_{V(S)}^{+}\rightarrow \mc{P}_{V(S)},\ \mc{P}_{V(S)}^{-}\rightarrow 0\, ,
\end{equation}
immediately eliminating all operators with insertions of $\mc{P}_{V(S)}^{-}$ from our consideration.  Similarly, operators of the form $\overline{N}_{V}\mc{P}_{V}^{+}\mc{P}_{S}^{+}N_{V}$ also vanish by projection.  The above discussion generalizes to any single nucleon $\mc{O}(a^{2})$ counterterm of the form 
\begin{equation}
	\overline{N}_V \mc{O} \mc{P}_V N_V \str [\mc{P}_S]\, ,
\end{equation}
and therefore we can parameterize the single nucleon counterterms relevant at NLO with the one operator given in Eq.~\eqref{eq:NMixCounterterm}.  It is important to note that this term, $\str [\mc{P}_S]$, respects chiral symmetry.

The mixed action two-nucleon (two-baryon, two-heavy meson) Lagrangian can be constructed in a similar fashion.  The mixed two-nucleon potential contains only two operators relevant at NLO,
\begin{align}
	\mc{V}_{NN}^\mathrm{Mix} &=C_{NN}^{a} \left( \overline{N}_{V}N_{V}\right)^{2}
	+C_{NN}^{b} \left( \overline{N}_{V}S^{\mu }N_{V}\right) ^{2}
	\notag\\&
	=D_{2a}^{({}^{1}S_{0})} \left( N_{V}^{T}P_{i}^{({}^{1}S_{0})}N_{V}\right)^{\dagger}
		\left(N_{V}^{T}P_{i}^{({}^{1}S_{0})}N_{V}\right)
	\nonumber\\&\ 
	+D_{2a}^{({}^{3}S_{1})} \left( N_{V}^{T}P_{i}^{({}^{3}S_{1})}N_{V}\right)^{\dagger}
		\left( N_{V}^{T}P_{i}^{({}^{3}S_{1})}N_{V}\right)\, ,
\end{align}
where $P_{i}^{({}j)}$ are the $S$-wave projection operators for channel--$j$ in the two nucleon system~\cite{Kaplan:1998sz}.  Through the order we are working, there are no counterterms for the other angular momentum projections because only the $S$-wave two-nucleon wavefunctions are non-vanishing at the origin, and so counterterms for the higher partial waves must also contain derivatives pushing the mixed action counterterms beyond the order we are working.

From the above construction, one observes that the leading $\mc{O}(a^{2})$ effects in $\mc{V}_{N}^\mathrm{Mix}$ and $\mc{V}_{NN}^\mathrm{Mix}$ appear to additively renormalize the sea quark mass dependent terms%
%
\begin{equation}
	\overline{N}_{V}N_{V}\mathrm{str}[ m_Q ]\, ,
\end{equation}
and 
\begin{equation}
	D_{2}^{\prime ({}j)}\left( N_{V}^{T}P_{i}^{({}j)}N_{V}\right) ^{\dagger}
	\left(N_{V}^{T}P_{i}^{({}j)}N_{V}\right) \mathrm{str}[B_0 m_{Q}]\, ,
\end{equation}
in the PQ$\chi$PT Lagrangians~\cite{Chen:2001yi,Beane:2002vq} and \cite{Beane:2002np} respectively.
However, it is important to stress that these lattice spacing dependent operators do not break chiral symmetry, and will differ from the sea quark mass dependent operators at higher orders.  These terms are allowed because the nucleon mass and the $NN$ interactions are not protected by chiral symmetry.

As discussed in Ref.~\cite{Chen:2006wf}, the $\mc{O}(a^{2})$ correction in $\mc{V}_{\phi}^\mathrm{Mix}$ only gives additive renormalization to the valence-sea meson masses while the subleading $\mc{O}(a^{2}m_{q})$
corrections only give additive renormalization to the LO meson decay
constant $f$ and the chiral condensate, $B_0$. Thus one can obtain the mixed action EFT
results at $\mc{O}(a^{2})$ from the PQ$\chi$PT results almost effortlessly.%
\footnote{This entire discussion can also be extended to heavy mesons as well as baryons with a heavy quark.  These fields also transform with $U(x)$ under chiral transformations and so the construction of the Lagrangian is very similar to that of the nucleon presented here.  The relevant partially quenched Lagrangians can be found in Ref.~\cite{Savage:2001jw} for the heavy mesons and in Refs.~\cite{Tiburzi:2004kd,Tiburzi:2004mv,Mehen:2006vv} for the baryons with heavy quarks.} 
We re-emphasize that this entire discussion and the prescription we present in Sec.~\ref{sec:MAprescription} is dependent upon these very benign mixed action lattice artifacts.  This simple behavior does not hold beyond the leading loop order except in special cases.  With these caveats in mind, we now provide the mixed action Lagrangian relevant for determining the chiral extrapolation formulae for all mixed action theories with chirally symmetric valence fermions.

%
%
\subsection{The Mixed Action Lagrangian \label{eq:MALagrangian}}
For no reason other than author bias, we present the Lagrangian in Minkowski space-time, despite the lattice theories being constructed in Euclidean space-time.  The LO mixed meson Lagrangian is%
\footnote{See Refs.~\cite{Chen:2005ab,Chen:2006wf} for our conventions labeling the meson fields and quark masses.} 
\begin{equation}\label{eq:LO_MA}
	\mc{L}_\phi^{(MA)} = \frac{f^2}{8} \str \left( \partial_\mu \S \partial^\mu \S^\dagger \right)
		+\frac{f^2B_0}{4} \str \left( m_Q \S^\dagger +\S m_Q^\dagger \right)
		+a^2 C_\mathrm{Mix} \str \left( T_3 \S T_3 \S^\dagger \right)
		+a^2 \mc{V}_{sea}\, .
\end{equation}
We have assumed that the sea-quark action is either $\mc{O}(a)$ improved, or has scaling violations (lattice spacing artifacts) beginning at $\mc{O}(a^2)$ or $\mc{O}(\a_s a^2)$.  This is not essential for our discussion, but as we will briefly discuss below, it simplifies the structure of the hairpin propagators such that to the order we are working, they are identical to those of PQ$\chi$PT~\cite{Bernard:1993sv,Sharpe:1997by,Sharpe:2000bc,Sharpe:2001fh}. We will nevertheless adopt this assumption below.

The LO Lagrangian, Eq.~\eqref{eq:LO_MA}, gives rise to the masses of the various pseudo-Goldstone mesons, with the LO masses for a meson composed of valence quarks, $v$, sea quarks $s$ or an admixture given by
\begin{align}\label{eq:LO_phiMass}
	m_{v_1 v_2}^2 &= B_0(m_{v_1} +m_{v_2})\, ,
	\nonumber\\
	\tilde{m}_{vs}^2 &= B_0(m_{v} +m_{s}) +a^2 \D_\mathrm{Mix}\, , 
	\nonumber\\
	\tilde{m}_{s_1 s_2}^2 &= B_0(m_{s_1} +m_{s_2}) +a^2 \D_{sea}\, ,
\end{align}
with
\begin{equation}
	a^2 \D_\mathrm{Mix} = a^2 \frac{16 C_\mathrm{Mix}}{f^2}\, ,
\end{equation}
and $a^2 \D_{sea}$ determined from $a^2 \mc{V}_{sea}$.  The LO Lagrangian also leads to the well known double-poles or hairpin propagators amongst the flavor diagonal mesons~\cite{Bernard:1993sv,Sharpe:1997by,Sharpe:2000bc,Sharpe:2001fh}.  The momentum space propagator between two flavor diagonal mesons of quark type $a$ and $b$ is given by
\begin{equation}\label{eq:etaProp}
	\mc{G}_{\eta_a \eta_b}(p^2) =
		\frac{i \e_a \d_{ab}}{p^2 -m_{\eta_a}^2 +i\e}
		- \frac{i}{N_f} \frac{\prod_{k=1}^{N_f}(p^2 -\tilde{m}_{k}^2 +i\e)}
			{(p^2 -m_{\eta_a}^2 +i\e)(p^2 -m_{\eta_b}^2 +i\e) \prod_{k^\prime=1}^{N_f-1}
				(p^2 -\tilde{m}_{k^\prime}^2 +i\e)},
\end{equation}
where
\begin{equation}
	\e_a = \left\{ \begin{array}{ll}
			+1& \text{for a = valence or sea quarks}\\
			-1 & \text{for a = ghost quarks}\,.
			\end{array} \right.
\end{equation}
In Eq.~\eqref{eq:etaProp}, $k$ runs over the flavor neutral states
($\phi_{jj}, \dots, \phi_{rr}$) and $k^\prime$ runs over the
mass eigenstates of the sea sector, including the additive lattice spacing mass corrections, Eq.~\eqref{eq:LO_phiMass}.  To help quantify the unitarity violating corrections arising from these double pole propagators, we have introduced partial quenching parameters~\cite{Chen:2006wf}, which are differences between the pole masses of the sea-sea and valence-valence mesons,
\begin{align}\label{eq:PQparameters}
	\tilde{\D}_{ju}^2 &\equiv \tilde{m}_{jj}^2 - m_{uu}^2
		= 2 B_0 (m_j- m_u) + a^2 \Delta_{sea} +\dots\, , \nonumber\\
	\tilde{\D}_{rs}^2 &\equiv \tilde{m}_{rr}^2 - m_{ss}^2
		= 2 B_0 (m_r- m_s) + a^2 \Delta_{sea} +\dots\, ,
\end{align}
where the dots denote higher order corrections to the meson masses.  Using these PQ parameters, one can rewrite the hairpin propagators in a particularly simple form allowing for a transparent identification of the unphysical unitarity violating contributions to correlation functions arising from the hairpin interactions~\cite{Chen:2005ab,Chen:2006wf}.

For discretization errors in the sea sector which begin at $\mc{O}(a)$, such as Wilson fermions~\cite{Wilson:1974sk}, there are two modifications we need to make to $\tilde{\D}_{ju}^2$ and $\tilde{\D}_{rs}^2$.  First, the lattice spacing corrections to the sea-sea meson mass begin at $\mc{O}(a)$.  Second, there are additional hairpin interactions whose coefficients depend upon the lattice spacing~\cite{Golterman:2005xa}.  This is not problematic to our prescription because these extra hairpins can be treated as an additional additive $\mc{O}(a^2)$ correction to our partial quenching parameters~\cite{Golterman:2005xa},
\begin{align}\label{eq:mod_DPQ}
	\tilde{\D}_{ju}^2 \longrightarrow&\ \tilde{m}_{jj}^2 -m_{uu}^2 + a^2 \g_{ss}\, N_s
	\nonumber\\& = 2 B_0 (m_j- m_u) + a W + a^2 \g_{ss}\, N_s\, ,
	\nonumber\\
	\tilde{\D}_{rs}^2 \longrightarrow&\ 2 B_0 (m_r- m_s) + a W + a^2 \g_{ss}\, N_s\, .
\end{align}
If we work consistently to $\mc{O}(a^2)$, we must include these extra hairpin effects, even though they are formally subleading to the $\mc{O}(a)$ term in the partial quenching parameters.  However, determining $\g_{ss}$ is difficult because it is an additive mass correction to the $\eta^\prime$ mass.  This shift in partial quenching parameters will also invariably cause a shift in the numerical value of unphysical lattice spacing dependent counterterms but this is accommodated without changing the structure of the extrapolation formulae.  We will generally assume that the sea fermion scaling violations begin at $\mc{O}(a^2)$ or higher as this is most relevant for lattice calculations of the present and future.
We will also commonly use the partial quenching parameters in the continuum limit to denote strictly differences in the sea and valence quark masses,
\begin{equation}
	\D_{ju}^2 = \tilde{\D}_{ju}^2 \Big|_{a=0} \quad,\quad
	\D_{rs}^2 = \tilde{\D}_{rs}^2 \Big|_{a=0}\, .
\end{equation}

%
%
\subsubsection*{\textbf{Mixed Action Single Baryon Lagrangian}}
The mixed action Lagrangian for the single nucleon (baryon) and interactions with the pions (mesons) is given by
\begin{align}\label{eq:LO_MA_Npi}
	\mc{L}_{N\phi}^{(MA)} =&\ i \left( \bar{\mc{B}} v\cdot D \mc{B} \right)
		+2 \a_M^{(PQ)} \left( \bar{\mc{B}} \mc{B} \mc{M}_+ \right)
		+2 \b_M^{(PQ)} \left( \bar{\mc{B}} \mc{M}_+ \mc{B} \right)
		+2 \s_M^{(PQ)} \left( \bar{\mc{B}} \mc{B} \right) \str ( \mc{M}_+ )
	\nonumber\\&\ 
		-\left( \bar{\mc{T}}^\mu \left[ i v\cdot D -\D \right] \mc{T}_\mu \right)
		+2 \g_M^{(PQ)} \left( \bar{\mc{T}}^\mu\, \mc{M}_+\, \mc{T}_\mu \right)
		-2 \s_M^{(PQ)} \left( \bar{\mc{T}}^\mu \mc{T}_\mu \right) \str \left( \mc{M}_+ \right)
	\nonumber\\&\ 
		+2 \a^{(PQ)} \left( \bar{\mc{B}} S^\mu \mc{B} \mc{A}_\mu \right)
		+2 \b^{(PQ)} \left( \bar{\mc{B}} S \cdot \mc{A} \mc{B} \right)
		+2\mc{H} \left( \bar{\mc{T}}^\nu\, S^\mu \mc{A}_\mu\, \mc{T}_\nu \right)
	\nonumber\\&\ 
		+\sqrt{\frac{3}{2}} \mc{C} \left( \bar{\mc{T}}^\mu\, \mc{A}_\mu\, B
			+\bar{B}\, \mc{A}_\mu\, \mc{T}^\mu \right)
		+a^2 C_a^{N} \left( \bar{\mc{B}} \mc{B} \right) 
		-a^2 C_a^T \left( \bar{\mc{T}}^\mu \mc{T}_\mu \right)\, .
\end{align}
Notice that there are only two new operators in the mixed action Lagrangian as compared to the partially quenched Lagrangian for either $SU(6|3)$~\cite{Chen:2001yi} or $SU(4|2)$~\cite{Beane:2002vq}.  If we were not projecting onto the valence sector of the theory or we were interested in working to higher orders, we would need additional lattice spacing dependent operators.  For example, there are $\mc{B}$-field operators similar to those with the mass spurion field $\mc{M}_+$ with chiral symmetry breaking $a^2$ spurions instead.  However, for extrapolation formulae of valence quantities, these operators all collapse into the form given in Eq.~\eqref{eq:LO_MA_Npi} at the order we are working.

The flavor structure and contractions of these fields, defined with the braces, $(\ )$, can be found in Ref.~\cite{Chen:2001yi} for the three flavor EFT and in Ref.~\cite{Beane:2002vq} for the two flavor EFT.  One often encounters formulae expressed with the more familiar $\chi$PT couplings instead of the PQ couplings, for example the $SU(4|2)$ couplings can be related to the $SU(2)$ couplings,
\begin{align}
	&g_A = \frac{2}{3}\a^{(PQ)} - \frac{1}{3}\b^{(PQ)},& 
	&g_1 = \frac{1}{3}\a^{(PQ)} + \frac{4}{3}\b^{(PQ)}&
	\nonumber\\
	&\mc{H} = g_{\D\D},& 
	&\mc{C} = -g_{\D N}&
\end{align}
The mass spurion field is given by
\begin{equation}
	\mc{M}_+ = \frac{1}{2} \left( \xi^\dagger m_Q \xi^\dagger + \xi m_Q^\dagger \xi \right)\, ,
\end{equation}
and the axial-meson field is
\begin{equation}
	\mc{A}_\mu = \frac{i}{2} \left( \xi \partial_\mu \xi^\dagger - \xi^\dagger \partial_\mu \xi \right)\, .
\end{equation}

%
%
\subsubsection*{\textbf{Mixed Action Two--Nucleon Lagrangian}}
Following the normalization and conventions of Ref.~\cite{Kaplan:1998sz}, the mixed action two-nucleon Lagrangian also involves simple modifications to the continuum two--nucleon Lagrangian~\cite{Weinberg:1990rz,Weinberg:1991um,Weinberg:1992yk,Kaplan:1998tg,Kaplan:1998we,Beane:2002nu} and is given by
\begin{align}
	\mc{L}_{NN}^\mathrm{Mix} =&  
	-C_{0}^{(j)} \left( N_{V}^{T}P_{i}^{(j)}N_{V}\right)^{\dagger}
		\left(N_{V}^{T}P_{i}^{(j)}N_{V}\right)
	+\frac{C_{2}^{(j)}}{8} \left( N_{V}^{T}P_{i}^{(j)}N_{V}\right)^{\dagger}
		\left(N_{V}^{T}P_{i}^{(j)} (\overleftrightarrow{\nabla})^2 N_{V}\right)
	\nonumber\\& 
	-\left( N_{V}^{T}P_{i}^{(j)}N_{V}\right)^{\dagger} \left(N_{V}^{T}P_{i}^{(j)}N_{V}\right) 
		\left[ D_{2B}^{(j)}\, \str(Bm_Q) + a^2 D_{2a}^{(j)} \right]
	\nonumber\\& 
	-D_{2A}^{(j)} \left( N_{V}^{T}P_{i}^{(j)}N_{V}\right)^{\dagger}
		\left(N_{V}^{T}P_{i}^{(j)}\, 2Bm_Q\, N_{V}\right)\, ,
\end{align}
where there is an implicit sum over $j= \left\{ {}^1 S_0, {}^3 S_1 \right\}$.  There are additionally operators with the $j={}^3 D_1$ projectors as well as the coupled ${}^3 S_1$--${}^3 D_1$ system necessary for understanding the deuteron, but as we discussed in Sec.~\ref{sec:Spurions} the lattice spacing counterterms for these channels are suppressed beyond the order we are considering.

\bigskip
In all of these MA Lagrangians, we see there are only very benign lattice spacing dependent operators.  This is crucial to the universal nature of the mixed action extrapolation formulae and a key component that allows us to construct our universal prescription.  Even though the detailed form of the Lagrangian at NLO depends upon the type of sea fermions employed, when we project onto the valence sector of the theory, all that is necessary to discuss the renormalization of valence quantities, we see this remarkable simplification of the relevant Lagrangian, as has been discussed in this section.  It is also clear that the simple form this Lagrangian takes depends crucially upon the chiral symmetry of the valence fermions, without which there would be several additional operators as is seen with the Wilson~\cite{Sharpe:1998xm,Rupak:2002sm,Bar:2003mh,Beane:2003xv,Tiburzi:2005vy} twisted mass~\cite{Munster:2003ba,Scorzato:2004da,Sharpe:2004ny,WalkerLoud:2005bt} and staggered~\cite{Lee:1999zxa,Aubin:2003mg,Sharpe:2004is,Bailey:2007iq} chiral Lagrangians.  We postpone a discussion of the class of observables we know to not follow our prescription until Sec.~\ref{sec:EDM}, where we use the neutron EDM as an example to highlight both the reasons this discussion fails to accurately describe this quantity as well as how to modify the prescription we provide in Sec.~\ref{sec:MAprescription} to account for this class of observables which is more sensitive to the sea-sector.

%
%
\subsection{Mixed Action Meson Operator as a Mixed Meson Mass}

To complete our discussion on MA Lagrangians at the one loop level, we prove that the mixed action meson operator functions exactly as the LO quark mass operator for a process with two valence-sea mesons, and $2N-2$ valence-valence mesons.  Let us consider the mixed lattice potential of Eq.~\eqref{eq:LO_MA}.  Making use of Eq.~\eqref{eq:T3} and the equality $\S \S^\dagger=1$, one can show that up to a constant this is equal to
\begin{align}
	a^2 C_\mathrm{Mix} \str \left( T_3 \S T_3 \Sigma^\dagger \right)
	&= 4a^2 C_\mathrm{Mix} \str \left( \mc{P}_S \S \mc{P}_S \Sigma^\dagger \right)
	\nonumber\\
	&= 4a^2 C_\mathrm{Mix} \sum_{N=0}^\infty \left( \frac{2 i}{f} \right)^{2N} \sum_{n=0}^{2N} 
	\frac{(-)^{n} \str( \mc{P}_S \phi^n \mc{P}_S \phi^{2N-n} )}{n! (2N-n)!}\, .
\end{align}
The only terms in the sum which contribute to vertices with two mixed valence-sea mesons come from either $n=0$ or $n=2N$, for which one can show
\begin{align}
	a^2 C_\mathrm{Mix} \str \left( T_3 \S T_3 \Sigma^\dagger \right)
	\longrightarrow&\ 8\, a^2 C_\mathrm{Mix} \sum_{N=0}^\infty \left( \frac{2 i}{f} \right)^{2N}
		\str \left( \frac{\mc{P}_S \phi^{2N}}{(2N)!} \right)
	\nonumber\\&\ 
	= 4\, a^2 C_\mathrm{Mix} \str \left[ \mc{P}_S \left( \S +\S^\dagger \right) \right]\, ,
\end{align}
and therefore the ratio of this mixed meson operator restricted to vertices involving two valence-sea mesons and $2N-2$ valence-valence%
\footnote{This also holds for vertices with $2N-2$ sea-sea mesons and two mixed valence-sea mesons.} 
mesons to the valence-sea quark mass contribution of the same vertices is
\begin{equation}\label{eq:C_Mix_Ratio}
	\frac{a^2 C_\mathrm{Mix} \str \left[ T_3 \S T_3 \S^\dagger \right]}
		{(f^2 /4) \str [ B_0m_Q(\S+\S^\dagger)]} \bigg|^{2\phi_{vs}}_{(2N-2)\phi_{vv}}
	= \frac{16\, a^2 C_\mathrm{Mix}}{f^2}
	= a^2 \D_\mathrm{Mix}\, .
\end{equation}
This is evident in previous determinations of mixed action extrapolation formulae~\cite{Chen:2005ab,O'Connell:2006sha,Bunton:2006va,Aubin:2006hg,Chen:2006wf,WalkerLoud:2006ub}.  Thus, the effects of this operator at one-loop act simply as a shift in the mixed valence-sea meson masses in all vertices and propagators.

%
%
\subsection{Mixed action EFT at one loop}

The symmetry structure of the underlying mixed lattice action determines the operators in the mixed action chiral Lagrangian.  The symmetries enjoyed by the valence fermions are different from those enjoyed by the sea fermions.  In the class of mixed action theories we consider here, the valence fermions only break chiral symmetry through the explicit quark mass term.  Therefore, in the meson Lagrangian at NLO, the purely valence spurions are identical to the spurions in continuum, unquenched chiral perturbation theory, and so the valence-valence sector of the NLO mixed action chiral Lagrangian is the Gasser-Leutwyler Lagrangian.  This is not the case for baryons which have LO lattice spacing operators as we have seen in Eq.~\eqref{eq:LO_MA_Npi}.  
The sea sector is different. At finite lattice spacing, the sea sector has enhanced sources of chiral symmetry violation --- for example, there are additional spurions associated with taste violation if the sea quarks are staggered, or in the case of a Wilson sea, the Wilson term violates chiral symmetry. Consequently, there are additional spurions in the sea sector. Of course, these spurions must involve the sea quarks and must vanish when the sea quark fields vanish.  

In this paper, we work consistently to NLO in the mixed action $\chi$PT power counting which we have defined in Eq.~\eqref{eq:power_counting}. For meson observables, the NLO operators in the Lagrangian are only used as counterterms; that is, at NLO one only computes at tree level with the NLO operators. Since the in/out states used in lattice simulations involve purely valence quarks, we can project the NLO operators onto the valence quark sector of the theory.  Consequently, all of the spurions which involve the sea quark fields vanish. Since the remaining spurions involve the valence quarks alone, we only encounter the symmetry structure of the valence quarks as far as the NLO operators are concerned. These spurions only depend on quark masses and the quark condensate itself, and so there can be no dependence on lattice discretization effects arising in this way.  The exception to this argument arises in the case of double trace operators in the NLO chiral Lagrangian; in these cases the valence and sea sectors interact in a flavor-disconnected manner, unlike the mixed operator in Eq.~\eqref{eq:VMix}.  If one trace involves a valence-valence spurion while the other involves a sea-sea spurion, then the trace over the sea may still contribute to a physical quantity, for example the meson masses and decay constants. Note that the valence-valence operators which occur in these double trace operators must be proportional to one of the two operators present in the LO chiral Lagrangian, Eq.~\eqref{eq:LO_MA}.  Thus, as it was argued and demonstrated in Ref.~\cite{Chen:2006wf}, for meson scattering processes, the dependence upon the sea quarks from these double trace operators can only involve a renormalization of the leading order quantities $f$ and $B_0$.  Both the explicit sea quark mass dependence and the explicit lattice spacing dependence are removed from the scattering processes expressed in terms of the lattice-physical parameters since they are eliminated in favor of the decay constants and meson masses which can simply be measured on the lattice.  When expressed in lattice-physical parameters, there can be no dependence upon the sea quark masses leading to unphysical PQ counterterms and similarly there can be no dependence upon an unphysical lattice-spacing counterterm.  This argument generalizes to all meson quantities which are protected by chiral symmetry.  For completeness, we summarize the discussion in Ref.~\cite{Chen:2006wf} here.

The NLO Lagrangian for mesons describing the valence and sea quark mass dependence is the Gasser-Leutwyler Lagrangian with traces replaced by supertraces:
\begin{align}
\mathcal{L}_{GL} =&\ L_1 \left[ \str \left( \partial_\mu \Sigma \partial^\mu \Sigma^\dagger \right) \right]^2
	+ L_2 \; \str \left( \partial_\mu \Sigma \partial_\nu \Sigma^\dagger \right) \str \left( \partial^\mu \Sigma 		\partial^\nu \Sigma^\dagger \right) 
	\nonumber\\&
	+ L_3 \; \str \left( \partial_\mu \S \partial^\mu \S^\dagger \partial_\nu \S \partial^\nu \S^\dagger 
		\right) 
	+ 2 B_0\, L_4 \; \str \left( \partial_\mu \S \partial^\mu \S^\dagger \right) 
		\str \left ( m_q \S^\dagger + \S m_q^\dagger \right) 
	\nonumber\\&
	+ 2 B_0\, L_5 \; \str \left[ \partial_\mu \S \partial^\mu \S^\dagger 
		\left( m_q \S^\dagger + \S m_q^\dagger \right) \right] 
	+ 4 B_0^2\, L_6 \; \left[ \str \left( m_q \S^\dagger + \S m_q^\dagger \right) \right]^2 
	\nonumber\\&
	+ 4 B_0^2\, L_7 \; \left[ \str \left( m_q^\dagger \S - \S^\dagger m_q \right) \right]^2
	+ 4 B_0^2\, L_8 \; \str \left( m_q \S^\dagger m_q \S^\dagger 
		+ \S m_q^\dagger \S m_q^\dagger \right).
\label{eq:LGL}
\end{align}
Having a concrete expression for the Lagrangian, we can show explicitly how the sea quark mass dependence disappears. The key is that when constructing NLO correlation functions of purely valence quarks, we can replace the mesonic matrix $\Phi$ in the NLO Lagrangian by a projected matrix
\begin{equation}
\Phi \rightarrow \mc{P}_V \Phi \mc{P}_V
\end{equation}
where $\mc{P}_V$ is the projector onto the valence subspace.  The only sea quark mass dependence comes from two operators
\begin{align}
\delta \mathcal{L}_{GL} =&\ 
	4 B_0\, L_4 \; \mathrm{str} \left( \partial_\mu \S P_V\, \partial^\mu \S^\dagger P_V \right) 
		\mathrm{str} (m_q) 
\nonumber\\& 
	+ 16 B_0^2\, L_6 \; \mathrm{str} \left(m_q \S^\dagger P_V + P_V\S m_q^\dagger\right)
		\mathrm{str} (m_q)\, ,
\end{align}
which leads to a renormalization of the LECs $f$ and $B_0$
\begin{align}
&f^2 \rightarrow f^2 + 32 L_4\, B_0\, \mathrm{str} (m_q),&
&f^2 B_0 \rightarrow f^2 B_0 + 64 L_6\, B_0^2\, \mathrm{str} (m_q).&
\end{align}
Since the parameters $f$ and $B_0$ are eliminated in lattice-physical parameters in favor of the measured decay constants and meson masses, we can remove the dependence on the sea quark masses by
working in lattice-physical parameters.  Analogously, we can remove all the explicit lattice spacing dependence.  The general MA Lagrangian involving valence-valence external states at $\mc{O}(\varepsilon_m^2 \varepsilon_a^2)$ can be reduced to the following form
\begin{align}
\delta \mc{L}_{MA} =&\ 
	a^2 \, L_{a^2}^{m_q}\, \str \left(m_q P_V \S^\dagger P_V + P_V\S P_V m_q^\dagger\right) 
		\str \left( g(P_S \S P_S)\, g^\prime(P_S \S^\dagger P_S) \right)
\nonumber\\ & 
	+a^2 L_{a^2}^{\partial}\, 
	\str \left( \partial_\mu \S P_V\, \partial^\mu \S^\dagger P_V \right) 
		\str \left(  f(P_S \S P_S)\, f^\prime(P_S \S^\dagger P_S) \right) \, +h.c.,
\end{align}
where the $f$'s and $g$'s are functions dependent upon the sea-quark lattice action.  These then lead to renormalizations of the LO constants,
\begin{align}
f^2 &\rightarrow f^2 
	+8a^2\, L_{a^2}^{\partial}\, \str \left( f(P_S 1 P_S)\, f^\prime(P_S 1 P_S) \right)\, , 
	\nonumber\\
f^2 B_0 &\rightarrow 
	f^2 B_0 
	+4a^2\, L_{a^2}^{m_q}\, \str \left( g(P_S 1 P_S) g^\prime(P_S 1 P_S) \right) ,
\end{align}
and just as with the sea quark mass dependence, expressing physical quantities in terms of the lattice-physical parameters removes any explicit dependence upon the lattice spacing.

Together, these results show that at NLO, the only counterterms entering into the extrapolation formulae for mesonic observables protected by chiral symmetry are the same as the continuum Gasser-Leutwyler counterterms entering at NLO.  This lack of unphysical counterterms is desirable from the point of view of chiral extrapolations, but it also has another consequence. Loop graphs in quantum field theories are frequently divergent; there must be a counterterm to absorb these divergences in a consistent field theory. Since there is no counterterm proportional to $a^2$ or the sea quark masses, loop graphs involving these quantities are constrained so that they have no divergence proportional to $a^2$ or the sea quark masses.  This further reduces the possible sources of sea quark or lattice spacing dependence. For example, mixed valence-sea meson masses have lattice spacing shifts, so there can be no divergence involving the valence-sea meson masses.

%
%
\subsubsection{Dependence upon sea quarks \label{sec:MANLO_finite}}
At NLO in the effective field theory expansion, mesons composed of one or two sea quarks only arise in loop graphs. In particular, the valence-sea mesons can propagate between vertices where they interact with valence-valence mesons; these interactions involve the LO chiral Lagrangian~\eqref{eq:LO_MA}. Because the mixing term, Eq.~\eqref{eq:VMix} is universal, these interaction vertices are the same for all discretization schemes provided LO chiral perturbation theory is applicable. The sea-sea mesons only arise
at NLO in hairpins.  Therefore, we see that our NLO extrapolation formulae only depend on the LO chiral Lagrangian to quadratic order in the sea-sea sector and the LO chiral Lagrangian (with the mixing term) in the valence-sea sector.  The mixed meson splitting has recently been computed for domain-wall fermions on the coarse MILC lattices~\cite{Orginos:2007tw} and was found to be
\begin{align}
	a^2 \D_\mathrm{Mix} &\simeq (314 \pm 4 \textrm{ MeV})^2
\end{align}
for $a=0.125$~fm~\cite{Orginos:2007tw}.  It has also been determined on the fine MILC lattices as well~\cite{Aubin:2008wk}.

Note that the impact of using different sea quark discretizations in our work comes only from the value of the LECs of the unphysical operators in the Lagrangian.  Therefore, the same NLO extrapolation formulae can be used to describe simulations with different sea quark discretizations, provided that the appropriate mass shifts are taken into account.  In the case of staggered sea quarks, the sea-sea mass splitting which occurs in the MA formulae is that of the taste-identity, which has been computed~\cite{Aubin:2004fs}, and for the coarse MILC lattices, is given by 
\begin{equation}
	a^2 \D_{sea} = a^2 \D_I \simeq (450 \textrm{ MeV})^2\, ,
\end{equation}
for $a \simeq 0.125$~fm.  These mass shifts can only appear through the hairpin interactions at this order.  These terms will generally be associated with \textit{unphysical} MA/PQ effects which give rise to the enhanced chiral logarithms as well as additional finite analytic dependence upon the sea-sea as well as valence-valence
meson masses (and their associated lattice spacing dependent mass corrections).

For heavy baryon and heavy meson observables, the leading loop corrections are typically non-analytic in the quark mass, for example the mass corrections are $\mc{O}(\varepsilon^3)$.  Therefore, the only counterterms needed to renormalize the NLO corrections to heavy baryon and heavy meson quantities will be those which appear in the LO Lagrangian, Eq.~\eqref{eq:LO_MA_Npi}.  The NLO loop contributions for these observables will typically involve all types of mesons, valence-valence, valence-sea and sea-sea.  Therefore, working to NLO, we only need to know the mass corrections to the valence-sea and sea-sea mesons.

%
%
\subsection{Prescription for Mixed Action Extrapolation Formulae \label{sec:MAprescription}}
We now have all the ingredients to construct a prescription to determine all MA extrapolation formulae at the one loop level from the corresponding PQ$\chi$PT expressions, for mixed actions with chirally symmetric valence fermions and any type of sea fermion.  This prescription is more useful if one expresses the extrapolation formulae with on-shell renormalization.  We also restate that this prescription is relevant for theories with chirally symmetric valence fermions and the hairpin structure of PQ$\chi$PT.  Given a PQ extrapolation formula make the replacements:

\bigskip
\noindent
\textit{\textbf{1.} meson and quark masses:} exchange the one-loop valence-valence meson masses with the lattice-physical meson masses, $m_{uu} \rightarrow m_\pi$, where $m_\pi$ is the lattice-physical pion mass (or appropriate meson mass for other valence quark flavors).  Replace tree-level meson masses (equivalently quark masses) with the lattice-physical pion mass at the appropriate value of the quark mass, with NLO adjustments as needed for consistency in the chiral expansion; for example $2B_0 m_u \rightarrow m_\pi^2 - \d m_\pi^2 (NLO)$; $2B_0 m_s \rightarrow 2m_K^2 -m_\pi^2 -2\d m_K^2(NLO) + \d m_\pi^2(NLO)$.

\bigskip
\noindent
\textit{\textbf{2.} decay constants:} for the LO decay constant $f \rightarrow f_\pi-\d f_\pi(NLO)$ where $f_\pi$ is the lattice-physical pion decay constant measured on the lattice and $\d f_\pi(NLO)$ is the one loop correction to this LO value which is entirely determined in terms of the lattice physical parameters.  Obviously for expressions which are already expressed in the on-shell renormalization, $f_{NLO} \rightarrow f_\pi$.  Equivalently, use $f_K$ or some linear combination of $f_\pi$ and $f_K$ with appropriate NLO adjustments.%
\footnote{This replacement also holds for all couplings which appear in a given formula.  For example, LHPC has determined the coupling $g_A$ which appears in the nucleon-pion Lagrangian.  For an extrapolation of the nucleon mass with the same lattice action, one should use this value of $g_A$ in the extrapolation formula.} 

\bigskip
\noindent
\textit{\textbf{3.} mixed mesons:} 
$m_{ju}^2 \rightarrow \tilde{m}_{ju}^2 = \frac{1}{2}m_{uu}^2 +\frac{1}{2}m_{jj}^2 +a^2 \D_\mathrm{Mix}$, for a meson composed of a valence and sea quarks $u$ and $j$.%
\footnote{There is no unique way to define the mixed meson mass renormalization; however, the different methods only differ at NLO and higher in the mixed meson mass, and therefore this difference will be NNLO or higher for all other quantities.  As an alternative, for the mixed ``kaon" mass, one could make the replacement, $\tilde{m}_{ru}^2 = \frac{1}{2}m_K^2 + \frac{1}{2}m_{jr}^2 + a^2\D_\mathrm{Mix}$ at the degenerate sea-valence quark mass point.} 

\bigskip
\noindent
\textit{\textbf{4.} sea-sea mesons:} $m_{jr}^2 \rightarrow \tilde{m}_{jr}^2 = m_{jr}^2 +a^2 \D_{sea}$ for a sea-sea meson composed of sea quark flavors $j$ and $r$ with the appropriate additive mass renormalization for a given sea quark discretization method.%
\footnote{For a Wilson sea, the mass correction will be linear in the lattice spacing, $\d m^2 = aW$.  For a clover-improved Wilson sea~\cite{Sheikholeslami:1985ij}, the mass correction is quadratic in the lattice spacing and is given in Ref.~\cite{Bar:2003mh}.  For a twisted mass sea at maximal twist~\cite{Frezzotti:2000nk}, it is the $\pi^\pm$ meson mass which enters this expression and the mass correction can be found in Ref.~\cite{Sharpe:2004ny}, see the Appendix for details.  For a staggered sea it is the taste-identity meson mass which enters at this expression, which has been measured on the coarse MILC lattices~\cite{Aubin:2004fs}, $a^2 \D_I \simeq (446 \textrm{ MeV})^2$.} 

\bigskip
\noindent
\textit{\textbf{5.} lattice spacing dependent counterterms/ higher dimensional operators:}  Add lattice spacing dependent counterterms (higher dimensional operators) when necessary.  Often, this can be determined by enforcing the renormalization-scale independence of a given observable.

%
%
\section{MA Extrapolation Formulae}\label{sec:appl}
To demonstrate the ease with which our prescription can be applied to the existing partially quenched literature, we determine the mixed action extrapolation formulae for several physical quantities which are currently of significant interest to the physics community.  We note that there are several mixed action EFT papers already in existence which provide further non-trivial examples of this prescription~\cite{Tiburzi:2005is,Chen:2005ab,Bunton:2006va,Aubin:2006hg,Chen:2006wf,Jiang:2007sn}.

%
%
\subsection{Neutron Electric Dipole Moment \label{sec:EDM}}
The neutron electric dipole moment (EDM) is of great interest both theoretically as well as experimentally.  A non-vanishing neutron EDM would be direct evidence of $CP$ violations which could stem from the $\bar{\theta}$-term in the QCD Lagrangian.  There have been several lattice calculations of the neutron EDM over the years~\cite{Aoki:1989rx,Guadagnoli:2002nm,Berruto:2005hg,Shintani:2005xg,Shintani:2006xr} with continually improving techniques and precision.  All the calculations to date and the foreseeable future require extrapolations to the physical point and given the  strong possibility of a mixed action lattice calculation of this quantity, it is very relevant to determine the mixed action extrapolation formula for the neutron EDM.

Furthermore, the neutron EDM is interesting because it is an example of a quantity which does not follow our prescription.  The reason for this is straightforward; the neutron EDM is directly proportional to the QCD $\bar{\theta}$-term and therefore is a quantity which is sensitive to the axial $U(1)$ chiral anomaly which so to speak ``lives" in the sea-sector.  This is simple to understand in PQ$\chi$PT.  Upon performing a chiral $U(1)_A$ rotation on the valence fermions, one must perform an equal rotation upon the ghost ``fermions".  In this manner, the change in the measure of the anti-commuting valence fields is exactly cancelled by a change in the measure of the commuting ghost fields, leaving the theory invariant.%
\footnote{This is equivalent to the discussion in Ref.~\cite{Aoki:1990ix}.} 
However, a chiral $U(1)_A$ rotation of the sea-fermions is connected to the desired $\bar{\theta}$-term of QCD, hence the abuse of language, ``the chiral-anomaly lives in the sea-sector."  

This has non-trivial consequences upon the structure of the extrapolation formula for the neutron EDM.  We can conclude that in the continuum limit, the neutron EDM must be proportional to the sea-quark masses (strictly speaking a product of the sea-quark masses), because if one of the quark masses were zero, the $\bar{\theta}$-term is non-physical and thus the EDM must vanish in this limit.  This rules out counterterms to the neutron EDM which are proportional to only the valence quark masses.  Away from the continuum limit, things are more involved and in fact we will need additional operators which we did not include in Eq.~\eqref{eq:LO_MA_Npi}.  We begin with the QCD Lagrangian including the $\bar{\theta}$-term,
\begin{equation}\label{eq:QCD_theta}
	\mc{L} = \bar{q} \left[ i\Dslash - m_q \right] q 
		-\frac{1}{4} F_{\mu\nu} F^{\mu\nu}
		+ \frac{g^2 \bar{\theta}}{32 \pi^2} F_{\mu\nu} \tilde{F}^{\mu\nu}\, .
\end{equation}
In the continuum limit, the theta-term can be rotated into the quark mass matrix with an axial $U(1)$ transformation and then mapped into the chiral Lagrangian.  However, at finite lattice spacing, this $U(1)_A$ transformation will also modify the irrelevant operators in the Symanzik action which break chiral symmetry, for example the chromo-magnetic term in the Symanzik Wilson Lagrangian will pick up a phase,
\begin{equation}
	\bar{q} \s_{\mu\nu} F^{\mu\nu}q 
		\longrightarrow
		\bar{q}_L e^{i \phi} \s_{\mu\nu} F^{\mu\nu} q_R
		+\bar{q}_R e^{-i \phi} \s_{\mu\nu} F^{\mu\nu} q_L\, ,
\end{equation}
with the flavor matrix $\phi = \textrm{ diag}(\phi_u, \phi_d, \dots )$, similar to the quark mass matrix, and $\bar{\theta}=~-\sum_j \phi_j$.  This will then give rise to lattice spacing dependent operators in the nucleon Lagrangian which contain this complex phase.  For example, the heavy baryon Lagrangian for the nucleon fields will have an operator~\cite{Beane:2003xv},
\begin{equation}\label{eq:Lag_N_Wilson}
	\mc{L}_W^{\bar{\theta}} \supset 2 \a_a \bar{N} aW_+^{\bar{\theta}} N\, ,
\end{equation}
with
\begin{equation}\label{eq:W_spurion}
	aW_+^{\bar{\theta}} = \frac{aW_0}{2} \left( \xi e^{-i\phi} \xi + \xi^\dagger e^{i\phi} \xi^\dagger \right)\, .
\end{equation}
The $aW_+$ field is parity even and therefore in the absence of the $\bar{\theta}$-term, only contains even numbers of pions.  However, with the complex phase present, this spurion field also contains pions of an odd number, and in particular can contribute to the neutron EDM in the one-loop graphs displayed for example in Fig.~1 of Ref.~\cite{O'Connell:2005un}, in place of the quark mass spurion of the nucleon Lagrangian.  With mixed action theories, the four quark operators of Eq.~\eqref{eq:L6} do not break chiral symmetry and are therefore invariant under the $U(1)_A$ transformation.  Therefore these mixed operators do not contribute to the neutron EDM.

In our construction of the mixed action chiral Lagrangian, we have not included certain operators which are important in the study of the neutron EDM or any other quantity sensitive to the chiral anomaly, which do not generally contribute to observables at this order.  They stem from four-quark operators constructed from sea-quarks and in this case which also break chiral symmetry.  To understand these operators, it is convenient to construct chiral symmetry breaking spurions of definite parity
\begin{equation}
	\mc{P}_{\chi,S}^{\pm,\bar{\theta}} 
		= \frac{1}{2} \left( \xi^\dagger \mc{P}_{S} e^{i \phi} \xi^\dagger 
			\pm \xi \mc{P}_{S} e^{-i \phi} \xi \right)\, .
\end{equation}
In particular, for an $\mc{O}(a)$ improved sea fermion action, there are two additional operators we should add to the Lagrangian which are important at this order in the presence of the $\bar{\theta}$-term,%
\footnote{For Wilson sea fermions, there will be similar operators but which only scale linearly in the lattice spacing.  These are the generalizations of Eq.~\eqref{eq:Lag_N_Wilson} to the mixed action theory.} 
\begin{equation}\label{eq:MA_N_new}
	\mc{L}_{N\phi}^{(MA)} \rightarrow 
	\mc{L}_{N\phi}^{(MA,\bar{\theta})}
	+\frac{2 a^2 \a_a}{N_s} \left( \bar{B} \mc{P}_{\chi,S}^{+,\bar{\theta}} B \right)
		\str ( \mc{P}_{\chi,S}^{+,\bar{\theta}} )
	+\frac{2 a^2 \b_a}{N_s} \left( \bar{B} B \mc{P}_{\chi,S}^{+,\bar{\theta}} \right)
		\str ( \mc{P}_{\chi,S}^{+,\bar{\theta}} )\, .
\end{equation}
We can immediately understand why these operators are not important in general.  In the absence of the $\bar{\theta}$-term, because they are parity even they only create an even number of pion fields, and therefore will only contribute to valence quantities beyond NLO.  The term with no pions from $\mc{P}_{\chi,S}^{+,\bar{\theta}}$ gives rise to a mass correction of baryons with at least one sea quark.  These are also higher order than we are working.  However, these operators will contribute to vertices in the one-loop graph contributing to the neutron EDM, replacing the $m_Q$ spurion insertions in Fig.~1 of Ref.~\cite{O'Connell:2005un}.

The LO contribution to the neutron EDM comes from a loop diagram in $\chi$PT~\cite{Crewther:1979pi}, with one interaction given by the LO HB$\chi$PT Lagrangian~\cite{Jenkins:1990jv,Jenkins:1991es,Jenkins:1991ne,Jenkins:1991ts}, or Eqs.~\eqref{eq:LO_MA_Npi} and \eqref{eq:MA_N_new} in the MA theory.  After performing the $U(1)_A$ rotation to remove the $\bar{\theta}$-term from the QCD Lagrangian, the quark mass term becomes (for $SU(4|2)$)
\begin{equation}
	m_Q^{(\bar{\theta})} = \textrm{diag}( m_u e^{i \phi_u}, m_d e^{i \phi_d}, m_j e^{i \phi_j}, m_l e^{i \phi_l}, m_u e^{i \phi_u}, m_d e^{i \phi_d} )\, ,
\end{equation}
with the operators in Eq.~\eqref{eq:MA_N_new} picking up similar phases, and thus provide new contributions to the loop graphs for the neutron EDM.  This is not the entire story however.  The values of the phases, $\phi_u, \phi_d$ \textit{etc.}, are determined by the vacuum stability of the pion potential, requiring there to be no single pion vertices, which in the partially quenched theory leads to the relations in the small angle limit~\cite{O'Connell:2005un}
\begin{equation}\label{eq:vacuumStability}
	m_u\phi_u = m_d\phi_d = m_j\phi_j = m_l\phi_l\, ,
\end{equation}
\begin{equation}
	\phi_u = \frac{-\bar{\theta}m_j m_l}{m_u(m_j+m_l)}\ ,\ 
	\phi_d = \frac{-\bar{\theta}m_j m_l}{m_d(m_j+m_l)}\ ,\ 
	\phi_j = \frac{-\bar{\theta}m_l}{(m_j+m_l)}\ ,\ 
	\phi_l = \frac{-\bar{\theta}m_j}{(m_j+m_l)}\, .
\end{equation}
At finite lattice spacing, there will be additional contributions to the sea-sea meson potential arising from chiral symmetry breaking operators which will also pick up phases under the above mentioned $U(1)_A$ rotation.  These will be the same operators which provide additive mass corrections to the sea-sea mesons and their contributions to the neutron EDM (and vacuum stability) can be easily accommodated with additive corrections to the sea-quark masses in Eq.~\eqref{eq:vacuumStability}, which will depend upon the particular lattice action used in the sea-sector.  For example, with Wilson fermions, the pions receive an $\mc{O}(a)$ mass shift.  Requiring there to be no single pion terms from the LO pion potential leads to
\begin{equation}\label{eq:mjWilson}
	m_{j,l}^W = m_{q_{j,l}} + a W_0/2B_0\, ,
\end{equation}
where $W$ is related to the chromo-magnetic condensate, which appears in Eq.~\eqref{eq:W_spurion} for example.  This is defined in a similar fashion to the chiral condensate, $B$.  For improved Wilson fermions, there will be a similar additive correction at $\mc{O}(a^2)$ which can be determined from the meson potential given in Ref.~\cite{Bar:2003mh} of a similar form,
\begin{equation}\label{eq:mjIW}
	m_{j,l}^{IW} = m_{q_{j,l}} +a^2 \tilde{W}/2B_0\, .
\end{equation}

The staggered sea quark masses are protected from multiplicative mass renormalization by the taste-5 chiral $U(1)_A$ symmetry of the staggered action.  However, the staggered meson potential is not invariant under a taste-singlet $U(1)_A$ rotation and therefore there will be additive shifts to the vacuum stability condition proportional to the taste-Identity meson mass splitting, similar to the additive corrections to the topological susceptibility with staggered fermions~\cite{Billeter:2004wx}.  This amounts to a correction of the sea quark mass of Eq.~\eqref{eq:vacuumStability} for staggered fermions of,
\begin{equation}\label{eq:mjStag}
	m_{j,l}^{stag.} = m_{q_{j,l}} + a^2 \D_{I}/2B_0\, .
\end{equation}

Putting this all together, and specializing to the case of degenerate sea quark masses and lattice spacing corrections which begin at $\mc{O}(a^2)$, the 
MA extrapolation formula for the neutron EDM determined with our modified prescription and Ref.~\cite{O'Connell:2005un} is then given by%
\footnote{Using our arguments above regarding the vanishing of the neutron EDM if one of the sea-quark masses is zero, allows us to uniquely determine two of the counterterms in Ref.~\cite{O'Connell:2005un}, those being $d=f=0$.  Also note the opposite sign convention of our Lagrangian, Eq.~\eqref{eq:LO_MA_Npi} as compared to that used in Ref.~\cite{O'Connell:2005un}.} 
\begin{align}\label{eq:EDM_MA}
d_N^{(PQ)} =& -\frac{e \bar{\theta}\, \hat{m}_{sea}}{(4 \pi f_\pi)^2} \left[ 
		4 F_\pi \ln \left( \frac{m_\pi^2}{\mu^2} \right)
		+4F_{ju} \ln \left( \frac{ \tilde{m}_{ju}^2}{\mu^2} \right) 
		+\frac{1}{2} c(\mu) \right]
	\nonumber\\&
	-\frac{e \bar{\theta}\, a^2}{(4 \pi f_\pi)^2} \left[
		4F^{a^2}_{ju} \ln \left( \frac{ \tilde{m}_{ju}^2}{\mu^2} \right)
		+\frac{1}{2}\tilde{c}_{a^2}(\mu) \right]\, .
\end{align}
In this equation, the sea quark mass, $\hat{m}_{sea}$ is given by Eq.~\eqref{eq:mjIW}, \eqref{eq:mjStag}, or the appropriate variation thereof for a given sea quark discretization method.  Furthermore, it is simple to accommodate non-degenerate sea quark masses as in PQQCD~\cite{O'Connell:2005un}, in which case $\hat{m}_{sea}$ is really proportional to the product of sea quark masses included in the chiral Lagrangian, including the appropriate additive mass renormalizations discussed above.  For the case of Wilson sea fermions, the only difference is that the second line of Eq.~\eqref{eq:EDM_MA} scales linearly in the lattice spacing, as opposed to the quadratic scaling given, and $\hat{m}_{sea}$ is given by Eq.~\eqref{eq:mjWilson}.  These contributions arise from the mixed action generalization of Eq.~\eqref{eq:Lag_N_Wilson}.

In Eq.~\eqref{eq:EDM_MA}, $F_\pi$ and $F_{ju}$ are combinations of the coefficients of the operators in Eq.~\eqref{eq:LO_MA_Npi} and can be found in Eq.~(30) of Ref.~\cite{O'Connell:2005un} and the new mixed action contribution has a coefficient
\begin{equation}
	F^{a^2}_{ju} = g_A \a_a \left( \frac{1}{3} +\frac{q_j +q_l}{2} \right)
		-g_1 \left[ \frac{\b_a}{3} -(q_j+q_l) \left( \frac{\a_a}{4}+\frac{\b_a}{2} \right) \right]\, .
\end{equation}
The sea quark electromagnetic charges are given by $q_j$ and $q_l$.  We see that even with these new considerations, the form of this extrapolation formula is still independent of the type of sea fermions used, provided the leading lattice spacing effects are $\mc{O}(a^2)$.  There will be similar modifications to the extrapolation formula relevant to all quantities which are sensitive to the chiral anomaly.

%
%
\subsection{Twist-2 matrix elements of the nucleon \label{sec:gA}}

Twist-2 matrix elements are related to moments of generalized parton distribution functions~\cite{Mueller:1998fv,Ji:1996ek,Radyushkin:1997ki}. Chiral perturbation theory was first applied to forward twist-2 matrix elements in \cite{Arndt:2001ye,Chen:2001eg,Chen:2001nb} then applied to off-forward matrix elements \cite{Chen:2001pva}. The quenched \cite{Chen:2001gra} and partially quenched \cite{Chen:2001yi} versions followed subsequently. Our approach can be used to convert the leading meson, single nucleon and multiple-nucleon twist-2 matrix elements to their mixed-action versions.  One of the more important twist-2 matrix elements is that related to the axial charge of the nucleon, which has recently been computed by the LHP Collaboration using a mixed action scheme~\cite{Edwards:2005ym}.  This is part of a more ambitious program to determine the structure of nucleons with lattice QCD and provides an important benchmark for the calculation of other twist-2 matrix elements~\cite{Edwards:2006qx,Hagler:2007xi}.  The nucleon axial charge can be calculated with the nucleon matrix element of the axial-vector current
\begin{equation}
	j_{\mu,5}^a =  \bar{q} \g_\mu \g_5 \t_a q\, ,
\end{equation}
which can be mapped into the heavy baryon chiral Lagrangian~\cite{Jenkins:1990jv,Jenkins:1991es}.  To determine the extrapolation formula for the calculation performed by LHPC, one can use the partially quenched formula worked out in Ref.~\cite{Beane:2002vq}, with the particular choice of extending the axial-charge matrix to 
\begin{equation}
	\bar{\t}^3 = \textrm{diag}(1,-1,0,0,1,-1)\, .
\end{equation}
Recently, the mixed action extrapolation formula for the neutron to proton axial matrix element was determined~\cite{Jiang:2007sn}, which in the isospin limit is equivalent to the proton-proton matrix element with the $\t_3$ current, and relevant for Ref.~\cite{Edwards:2005ym}.  In the mixed action EFT, the extrapolation formula is given at NLO by
\begin{equation}
	\langle\, N(p)\, |\, {}^{(MA)} j^3_{\mu,5}\, |\, N(p)\, \rangle = 2 \bar{u}_p\, S_\mu\, u_p
		\left[ {}^{(MA)}\Gamma_{pp} +{}^{(MA)}c_{pp} \right]\, ,
\end{equation}
where
\begin{align}\label{eq:Gpp_MA}
{}^{(MA)}\Gamma_{pp} =&\ g_A 
	-\frac{1}{(4\pi f_\pi)^2} \bigg[ 
	L(\tilde{m}_{ju}) \frac{1}{6} \left( 12 g_A + 24 g_A^3 +16 g_A^2 g_1 +17g_A g_1^2 +g_1^3 \right)
	\nonumber\\& 
	-L(m_\pi) \frac{g_1}{6} \left( 16g_A^2+17g_A g_1 +g_1^2 \right)
	+2 g_A(g_A+g_1)^2\, \tilde{\D}_{ju}^2 \frac{\partial}{\partial m_\pi^2} L(m_\pi)
	\nonumber\\&
	-\frac{4}{9}(4g_A +g_1)g_{\D N}^2 K(m_\pi,\D,\mu)
	-\frac{4}{9}(4g_A -g_1)g_{\D N}^2 K(\tilde{m}_{ju},\D,\mu)
	\nonumber\\&
	+2g_{\D N}^2 \left(g_A +\frac{10}{27} g_{\D\D} \right)\, J(m_\pi,\D,\mu)
	+2g_{\D N}^2 \left( g_A +\frac{20}{81} g_{\D\D} \right)\, J(\tilde{m}_{ju},\D,\mu)
	\bigg]\, ,
\end{align}
and 
\begin{equation}\label{eq:cpp_MA}
	{}^{(MA)}c_{pp} = m_\pi^2 C_m + \D_{ju}^2 C_m^{(PQ)} +a^2 C_a\, ,
\end{equation}
is given by local counter terms.  Our formula is in agreement with that in Ref.~\cite{Jiang:2007sn}, however as we will explain shortly, we have a slight disagreement with the analysis presented in Ref.~\cite{Jiang:2007sn} in the estimation of the size of the lattice spacing dependent corrections.  Before discussing the relevance of this formula to the LHPC calculation, first we contrast this formula with the continuum $\chi$PT formula which was used to perform the chiral extrapolation in Ref.~\cite{Edwards:2005ym},
\begin{align}\label{eq:Gpp_ChPT}
\Gamma_{pp} =&\ g_A
	-\frac{1}{(4\pi f_\pi)^2} \bigg\{
		2(g_A +2g_A^3)\, L(m_\pi)
	\nonumber\\&
	+\frac{4g_{\D N}^2}{81} (81g_A +25 g_{\D\D})\, J(m_\pi,\D,\mu)
	-\frac{32}{9}g_A g_{\D N}^2\, K(m_\pi,\D,\mu) \bigg\}\, ,
\end{align}	
and
\begin{equation}\label{eq:cpp_ChPT}
	c_{pp} = m_\pi^2 C_m\, .
\end{equation}
In both the $\chi$PT and MA formulae, $L$, $J$ and $K$ are chiral logarithm functions defined in the literature, with 
\begin{equation}
	L(m) = m^2 \ln \left( \frac{m^2}{\mu^2} \right)\, ,
\end{equation}
and $J$ and $K$ can be found for example in Eqs~(62) and (69) or Ref.~\cite{Beane:2002vq} respectively, and their finite volume equivalents in Ref.~\cite{Beane:2004rf}.  There are two important distinctions between the MA and $\chi$PT formulae.  The MA extrapolation formula has two more counterterms than the continuum formula, as seen by Eqs.~\eqref{eq:cpp_MA} and \eqref{eq:cpp_ChPT}, however with the tuning used by LHPC, $\D_{ju}\simeq0$ and therefore they require only one more counterterm.  The lattice spacing dependent counterterm in Eq.~\eqref{eq:cpp_MA} is required by scale invariance and also simply follows from the spurion analysis presented in Sec.~\ref{sec:Spurions} or from the arguments in Ref.~\cite{Jiang:2007sn}.  Additionally, the loop corrections in the MA formula depend upon two different pion-nucleon axial couplings, $g_A$ and $g_1$ as opposed to the one coupling, $g_A$ in the continuum $\chi$PT formula, Eqs.~\eqref{eq:Gpp_MA} and \eqref{eq:Gpp_ChPT}.  The MA formula also depends upon the mixed valence-sea mesons as well as the taste-identity staggered pion mass.  However, the staggered taste splittings are well known~\cite{Aubin:2004fs} and the mixed meson mass renormalization has recently been calculated and is also known well~\cite{Orginos:2007tw}.  Therefore the MA extrapolation of the nucleon axial charge calculated by LHPC requires the determination of two additional unknown, unphysical terms as compared to the continuum $\chi$PT formula.

Given the presence of these two unphysical LECs, and the limited amount of mass points in the LHPC calculation, one would like to estimate the size of the corrections to the continuum extrapolation formula due to the mixed action artifacts, to determine the impact these unphysical effects have in the extraction of the nucleon axial charge.  This was taken up in Ref.~\cite{Jiang:2007sn}, and we do not repeat the analysis here, but we highlight a point of disagreement we have regarding the size of the lattice spacing dependent corrections.  In Ref.~\cite{Jiang:2007sn}, the lattice spacing counterterm, $C_a$ of Eq.~\eqref{eq:cpp_MA}, was varied in an uncorrelated fashion with the partially quenched pion-nucleon coupling, $g_1$ and the mixed meson mass renormalization, which was not known at the time.  This lead to a predicted error band of $\mc{O}(200\%)$ of the value of the nucleon axial charge, $g_A^N$ as measured at $m_\pi \sim 350$~MeV by LHPC (Table~1 of Ref.~\cite{Edwards:2005ym}) which was almost entirely due to the mixed action lattice spacing dependent corrections, see Fig.~3 of Ref.~\cite{Jiang:2007sn}.  This signals either a breakdown in the mixed action expansion for this observable or an overestimate of the errors in Ref.~\cite{Jiang:2007sn}.  Given the quality of the lattice results and extrapolation performed by LHPC~\cite{Edwards:2005ym} and the small corrections the mixed action effects contribute to other quantities~\cite{Chen:2005ab,Chen:2006wf} we find the latter to be the more plausible explanation.

At a fixed lattice spacing, by performing the chiral extrapolation there is no way to distinguish between the lattice spacing dependent counterterm, $C_a$, the pion-mass independent contributions from ${}^{(MA)}\G_{pp}$  and the LO contribution to the axial charge.  Therefore, when LHPC performed their chiral extrapolation and determined a best fit value for chiral Lagrangian parameter $g_A$, they were actually determining the linear combination
\begin{equation}
	\tilde{g}_A = g_A + a^2 \tilde{C}_a\, 
\end{equation}
where $\tilde{C}_a$ is a linear combination of the lattice spacing dependent counterterm, $C_a$ and loop contributions to ${}^{(MA)}\Gamma_{pp}$ which do not vanish in the chiral limit and are proportional to either $a^2 \D_I$ from the hairpin contribution or $a^2 \D_\mathrm{Mix}$ from the mixed meson mass contributions.  If we then assume a perturbative expansion (which breaks down for light enough pion masses, as the hairpin interaction diverges in the chiral limit~\cite{Beane:2002vq,Jiang:2007sn}), the uncertainty in this extracted parameter is a much better estimate of the size of the lattice spacing artifacts in LHPC's calculation of the nucleon axial charge, and this is already contained in the error band presented in Figure~1 of Ref.~\cite{Edwards:2005ym}.  In support of this estimation, it has been found with other quantities, that the general size of the mixed action artifacts with the LHPC mixed action scheme~\cite{Renner:2004ck,Edwards:2005kw} have been at the 1--5\% level~\cite{Chen:2005ab,Chen:2006wf}.  A full analysis of the mixed action corrections will involve the determination of the coupling $g_1$ as well as making use of both the staggered taste-identity pion mass splitting~\cite{Aubin:2004fs} and the recently determined mixed meson mass renormalization~\cite{Orginos:2007tw} which is beyond the scope of this work.  We finally note that it was show in Ref.~\cite{Jiang:2007sn} that it is the uncertainty in the partially quenched parameter $g_1$ which dominates the uncertainty in the mixed action corrections to the nucleon axial charge.

\bigskip
More important to the LHPC program of calculating the structure of the nucleon~\cite{Edwards:2006qx,Hagler:2007xi}, one can use our prescription, the lattice spacing mass shifts calculated in Refs.~\cite{Aubin:2004fs,Orginos:2007tw} and the extensive use of the work of Detmold and Lin in Ref.~\cite{Detmold:2005pt}, in which the partially quenched extrapolation formulae for all the forward twist-2 matrix elements has been determined to NLO in both finite and infinite volume, to determine the corresponding mixed action formulae.%
\footnote{We note that the iso-vector twist-2 matrix elements were first determined in Refs.~\cite{Chen:2001yi,Beane:2002vq} for $SU(6|3)$ and $SU(4|2)$ partially quenched theories respectively.} 
Using the spurion analysis presented in Sec.~\ref{sec:Spurions}, one can show that each twist-2 matrix element will have its own lattice spacing dependent counterterms which can be treated as $a^2 C_a^{(n)}$ at NLO, as in Eq.~\eqref{eq:cpp_MA}.  As with the nucleon axial charge, the partially quenched pion--nucleon coupling $g_1$ will likely play the largest role in the mixed action corrections because it is precisely the linear combination $(g_A+g_1)^2$ which is the coefficient of the hairpin interaction contributions to all the twist-2 matrix elements.  Therefore it will be important to determine this coupling to have good control of the mixed action lattice artifacts.

%
%
\subsection{$NN$ Scattering \label{sec:NN}}

One of the greatest challenges facing the nuclear physics community is to
determine the properties of nuclei from QCD. Given the success of lattice
QCD with the meson and single nucleon sector calculations, it is natural to use
lattice QCD to study nuclear systems. One of the complications however, is that nuclear physics is a finely tuned system. The deuteron binding energy for example $B_d \simeq 2.24$~MeV, is much
smaller than the scale set by pion physics. This makes it quite formidable 
to extract the deuteron binding energy in a lattice calculation out of the approximately 2 GeV rest mass of the proton and neutron.  It turns out that by using heavier than physical light quark
masses (corresponding to $m_\pi \sim 300$~MeV), the two nucleon
scattering lengths, which can be related to binding energy, tend to
be a natural size and measurable with a box of $L\sim2.5$~fm per side~\cite{Beane:2006mx}.%
\footnote{There is some subtlety here.  When determining the two-particle energy levels from the spatial volume dependence of the four-point correlation function, commonly referred to as L\"{u}scher's method~\cite{Huang:1957im,Hamber:1983vu,Luscher:1986pf,Luscher:1990ux}, it is not the scattering length which determines the required size of the volume for the method to be applicable, but the effective range, generally set by the inverse pion mass in QCD.  Therefore, even with unnaturally large scattering lengths as with the two-nucleon system at the physical quark masses, one can determine the infinite volume scattering parameters from the two particle interaction energy even when the scattering length is much larger than the finite spatial extent of the lattice~\cite{Beane:2003da}.  A smaller scattering length in the ${}^3 S_1$ channel is indicative of a larger binding energy for the deuteron making it easier to determine on the lattice.  Also, the inverse box size, $L^{-1}$ determines the splitting in energy levels of the two-particle system which is important for having well separated eigenstates.  And lastly, one needs to make sure the exponentially suppressed volume modifications, which generically scale as $e^{-m_\pi L}$ are under control in the two-nucleon system~\cite{Sato:2007ms}, which is not as straight forward as with the two pion system~\cite{Bedaque:2006yi}.} 
To further extract the physical scattering lengths, extrapolations to the continuum infinite volume limit and the physical quark masses are required.

Recently, the NPLQCD Collaboration has determined the nucleon-nucleon
scattering lengths in the ${}^{1}S_{0}$ and ${}^{3}S_{1}$ channels using a
MA lattice calculation with domain-wall valence quarks and the asqtad
improved staggered MILC gauge configurations~\cite{Beane:2006mx}. It is
therefore of considerable interest to understand how the MA artifacts
pollute the correlation functions. In Ref.~\cite{Beane:2002nu}, Beane and
Savage have developed the partially quenched version of the two nucleon systems based on
BBSvK power counting~\cite{Beane:2001bc}. In BBSvK, the $^{1}S_{0}$ channel follows
the KSW power counting~\cite{Kaplan:1998tg,Kaplan:1998we} with one pion exchange entering as a perturbation at NLO while the $^{3}S_{1}$ channel follows Weinberg's power
counting~\cite{Weinberg:1990rz,Weinberg:1991um,Weinberg:1992yk} with one pion exchange entering non-perturbatively and being resumed to the LO one pion exchange potential.  The PQ effects
($\D_{ju}\neq 0$) in the two-nucleon system arise from several important sources: the one pion exchange diagram also includes a hairpin interaction modifying the long-distance part of the potential~\cite{Beane:2002nu}, the PQ corrections to the masses and couplings of the particles; $g_A$, $f_\pi$, $m_\pi$ and $m_N$, as well as two new PQ $NN$ couplings, $D_{2B}^{({}^{1}S_{0})}$ and $D_{2B}^{({}^{3}S_{1})}$.

Using the prescription described above, we can easily incorporate the MA effects in two nucleon quantities once the corresponding PQ effects are known. Now we go through the list of PQ effects listed above and modify them to include the MA effects. The one pion exchange potential (OPE) is%
\footnote{The hairpin modification to Eq.~\eqref{eq:VNN_MA} appears different to that in Refs.~\cite{Beane:2002nu,Beane:2002np}, however this is simply a different convention for labeling the pion--nucleon couplings.  Our convention is consistent with Ref.~\cite{Beane:2002vq}.} 
\begin{equation}\label{eq:VNN_MA}
V_{OPE}^{MA}(r) = \frac{1}{8\pi f_\pi^2}
	\vec{\s}_{1}\cdot \vec{\nabla}\,
	\vec{\s}_{2}\cdot \vec{\nabla}\left[ \,
		g_A^2\, \frac{\vec{\tau}_{1}\cdot \vec{\tau}_{2}}{r} 
		-(g_A+g_1)^2\, \frac{\tilde{\D}_{ju}^{2}}{2m_{\pi }}
	\right] e^{-m_{\pi }r}\, .
\end{equation}
in the ${}^1 S_0$ channel.  The only modification from the partially quenched potential determined in Ref.~\cite{Beane:2002nu} is $\D_{ju}\rightarrow \tilde{\D}_{ju}$.%
\footnote{Recall that $\D_{ju}=\tilde{\D}_{ju}|_{a=0}$.} 
The formula for the pion decay constant has been worked out in several places and we list it here for convenience,
\begin{equation}
	f_\pi = f \left[ \ 1
		-\frac{2\tilde{m}_{ju}^2}{(4\pi f)^2}
			\ln \left( \frac{\tilde{m}_{ju}^{2}}{\mu^2} \right)
		+2 \ell_{4}(\mu) \frac{m_{\pi}^{2}}{f^2} 
		+\ell_{f}^{(PQ)}(\mu) \frac{\D_{ju}^{2}}{f^2}
		+\ell_{a}^{(MA)}(\mu) \frac{a^{2}}{f^2}
		\right] \ .
\end{equation}%
The MA formula for the nucleon mass can be found in Ref.~\cite{Tiburzi:2005is} and $g_A$, which we showed in the previous section, can also be found in Ref.~\cite{Jiang:2007sn}.  The $NN$ counterterms, following the spurion analysis presented in Sec.~\ref{sec:Spurions}, should be replaced by
\begin{equation}
D_{2}^{({}j)}m_{\pi }^{2}\rightarrow 
	D_{2}^{({}j)}m_{\pi }^{2}
	+\D_{ju}^{2}D_{2B}^{({}j)}
	+a^{2}\,D_{2a}^{({}j)}\, .
\end{equation}
Making use of the work in Ref.~\cite{Beane:2002np}, we then find that in a MA theory with Ginsparg-Wilson valence fermions, the ${}^{1}S_{0}$ scattering length and effective range extrapolation formula are
\begin{align}
\frac{1}{a^{({}^{1}S_{0})}} =&\ \gamma 
	-\frac{M_{N}}{4\pi }(\mu -\gamma)^{2}\,D_{2}^{({}^{1}S_{0})}(\mu )\,m_{\pi }^{2}
	\nonumber\\&\ 
	+\frac{g_{A}^{2}M_{N}}{8\pi f_{\pi }^{2}}\left[ 
		m_{\pi }^{2}\,\ln \left( \frac{\mu }{m_{\pi }}\right) 
		+(m_{\pi }^{2}-\gamma )^{2}-(\mu -\gamma )^{2}
	\right]  
	\nonumber\\&\ 
	-\left( \D_{ju}^{2}D_{2B}^{({}^{1}S_{0})}(\mu) 
		+a^{2}\,D_{2a}^{({}^{1}S_{0})}(\mu )
	\right) \frac{M_{N}}{4\pi }(\mu -\gamma)^{2} 
	\nonumber\\&\ 
	+\tilde{\D}_{ju}^{2}\frac{(g_A+g_1)^{2}M_{N}}{8\pi f_{\pi }^{2}}\left[ \ln
\left( \frac{\mu }{m_{\pi }}\right) +\frac{1}{2}-\frac{\gamma }{m_{\pi }}%
\right]\, ,
\end{align}
and
\begin{align}
r^{({}^{1}S_{0})} =&\ 
	\frac{M_{N}}{2\pi }(\mu -\gamma )^{2}\,C_{2}(\mu )
	+\frac{g_{A}^{2}M_{N}}{12\pi f_{\pi }^{2}}\left( 3-8\frac{\gamma }{m_{\pi }}
		+6\frac{\gamma ^{2}}{m_{\pi }^{2}}\right)
	\nonumber\\&\ 
	+\frac{\tilde{\D}_{ju}^{2}}{m_{\pi }^{2}}\frac{(g_A+g_1)^2 M_{N}}{6\pi f_{\pi}^{2}}
		\left( 2\frac{\gamma }{m_{\pi }}-3\frac{\gamma ^{2}}{m_{\pi }^{2}}
		\right)\, ,
\end{align}
where $\g$ is a $\mu$-independent linear combination of $\mu$ and $C_0^{({}^1 S_0)}(\mu)$, the LO $NN$ interaction.  Compared to the PQ case, the ${}^{1}S_{0}$ scattering length has one new lattice spacing dependent counter term at this order, $D_{2a}^{({}^{1}S_{0})}(\mu )$ while the effective range does not depend upon any new counterterms.  The results in $^{3}S_{1}$ channel unfortunately do not have an analytic form because it requires solving the $^{3}S_{1}$-$^{3}D_{1}$ coupled Schrodinger equation.  However, the difference in the PQ and MA potentials follows the description given above, and can be determined from Ref.~\cite{Beane:2002np} with our prescription.

Lastly, we comment that this discussion naturally extends to the hyperon-nuclear interactions as well, for which the partially quenched theory has been developed in Ref.~\cite{Beane:2003yx}.  This is also very relevant as the first lattice study of the hyperon-nucleon interaction has recently been performed in a MA scheme as well~\cite{Beane:2006gf}.

%
%
\section{Discussion}\label{sec:concl}
In this work, we have proven that the new leading order meson operator allowed by mixed action theories, which is independent of the mixed lattice action, functions exactly as a meson mass operator for all vertices with $(2N-2)$ valence-valence mesons and 2 mixed valence-sea mesons.  This proof, combined with the very nice features of mixed action effective field theories with chirally symmetric valence fermions~\cite{Chen:2006wf} has allowed us to construct a prescription to determine many mixed action extrapolation formulae for valence quantities, through the leading loop order given the corresponding formulae determined in partially quenched chiral perturbation theory. Our prescription works immediately for quantities which do not depend on the $\bar \theta$ term; we have further assumed that the mixed action theory has the same hairpin structure as the partially quenched theory, and that the valence quarks are chiral. In the case of a mixed action theory with more involved hairpin structure than the partially quenched theory, a more general prescription is undoubtedly possible. We have used the neutron EDM as an example which requires a modification of our prescription as it depends critically upon the $\theta$ term, and have discussed how to modify our prescription for this quantity. With the recently measured mixed meson mass renormalization~\cite{Orginos:2007tw} and the well known staggered meson mass taste splittings~\cite{Aubin:2004fs}, these mixed action extrapolation formulae can readably be applied to a host of physical quantities covering a broad range of hadronic physics: pion and kaon physics~\cite{Bernard:1993sv,Golterman:1997st,Sharpe:1997by,Sharpe:2000bc,Sharpe:2001fh,Lin:2003tn,Chen:2005ab,Bunton:2006va,Aubin:2006hg,Chen:2006wf}, baryon observables~\cite{Chen:2001yi,Beane:2002vq,WalkerLoud:2004hf,Tiburzi:2004rh,Tiburzi:2005na,Jiang:2007sn}, heavy meson observables~\cite{Savage:2001jw}, heavy hadron observables~\cite{Tiburzi:2004kd,Tiburzi:2004mv,Mehen:2006vv}, parity violation~\cite{Beane:2002ca,O'Connell:2005un}, electromagnetic properties and transition matrix elements~\cite{Arndt:2003ww,Arndt:2003we,Arndt:2003vd,Detmold:2006vu}, structure functions~\cite{Detmold:2005pt,Chen:2006gg}, two nucleon systems~\cite{Beane:2002nu,Beane:2002np}, hyper-nuclear systems~\cite{Beane:2003yx} and constraints on beyond the standard model physics from hadronic contributions of $\D b =2$ and $\D c =2$ observables~\cite{Detmold:2006gh}.

%
%
\begin{acknowledgments}
We would like to acknowledge Fu-Jiun Jiang, Kostas Orginos, Gautam Rupak, Steve Sharpe and Brian Tiburzi for useful correspondence and conversations.  AWL would like to acknowledge the hospitality of the Department of Physics and Center for Theoretical Sciences at the National Taiwan University where the idea for this research was born out.  JWC is supported by the National Science Council of R.O.C.. DOC is supported in
part by the U.S. DOE under the grant DE-FG03-9ER40701.  AWL is supported
under the U.S. DOE grant DE-FG02-93ER-40762.
\end{acknowledgments}

%
%
\appendix
\section{Chiral valence fermions on twisted mass sea fermions \label{app:MA_tm}}

Here we present a few technical details of the mixed action chiral Lagrangian corresponding to chiral valence fermions and twisted mass sea fermions alluded to in the main text.  We focus on the chiral theory for two degenerate valence and ghost flavors as well as a degenerate pair of twisted sea flavors.  We will demonstrate that this theory behaves \textit{nicely} in the following sense; the $\mc{O}(a^2)$ mixed valence-sea fermion mass shift is given entirely by the $C_\textrm{Mix}$-term at maximal twist; the hairpin structure of the theory is the same as in partially quenched $\chi$PT at maximal twist to the order we are working.  These two facts follow naturally from the properties of graded algebras and the behavior of two flavor twisted mass QCD.  We first summarize the discussion for twisted mass $\chi$PT which is relevant  to our discussion.

The Euclidean Symanzik Lagrangian for twisted mass lattice QCD~\cite{Frezzotti:2000nk} is~\cite{Sharpe:2004ps}
\begin{equation}
\mc{L}_{eff} = \mc{L}_{glue} 
	+ \bar{\psi}\, \Big[ \Dslash + m +i \g_5 \t_3 \mu \Big]\,  \psi
	+ b_1 a\, \bar{\psi}\, i \s_{\mu\nu} F_{\mu\nu}\, \psi\, ,
\end{equation}
where $\mc{L}$ is the gluon action, $m = Z_m (m_0 - \tilde{m}_c )/a$ is the standard quark mass and $\mu = Z_\mu \mu_0 / a$ is the twisted quark mass.  From here one can construct the low energy chiral Lagrangian which is given at LO by~\cite{Munster:2003ba,Scorzato:2004da,Sharpe:2004ny}
\begin{equation}
\mc{L}_\chi = \frac{f^2}{8} \tr \left( \partial_\mu \S \partial_\mu \S^\dagger \right)
	-\frac{f^2}{8} \tr \left( \chi^\dagger \S + \S^\dagger \chi \right)
	-\frac{f^2}{8} \tr \left( A^\dagger \S + \S^\dagger A \right)\, ,
\end{equation}
where 
\begin{align}
	&\chi = 2B_0 (m + i\t_3 \mu ) \equiv \hat{m} + i\t_3 \hat{\mu}&
	&\text{and}&
	&A = 2W_0 a \equiv \hat{a}\, .&
\end{align}
The first observation to make is one can define a shifted spurion field such that the Lagrangian takes its continuum form, with
\begin{equation}
\chi^\prime = \chi + A
	\equiv M^\prime e^{i \w_0 \t_3}\, ,
\end{equation}
where $M^\prime = \sqrt{(\hat{m} + \hat{a})^2 + \hat{\mu}^2}$.  The second observation is that because of the twisted mass term, the vacuum will no longer be aligned with the identity, which can be determined by minimizing the vacuum energy.  One finds
\begin{equation}
\S_0 \equiv \langle 0 | \S | 0 \rangle = e^{i\w_0 \t_3}\, ,
\end{equation}
such that the physical fields are given by an axial rotation from $\S$,
\begin{equation}
	\S = e^{i\w_0 \t_3/2} \S_{phys} e^{i\w_0 \t_3/2}\, .
\end{equation}
Expanding the LO and NLO Lagrangians about the vacuum, one then finds the automatic $\mc{O}(a)$ improvement of physical observables~\cite{Munster:2003ba,Scorzato:2004da,Sharpe:2004ny}.  We now extend this analysis to the mixed action theory.  Working in the isospin limit of the valence and sea sectors, the mixed action Lagrangian including the leading $\mc{O}(a^2)$ operators is
\begin{align}\label{eq:MA_tm_Lag}
 \mc{L} =&\ \frac{f^2}{8} \str \left( \partial_\mu \S \partial_\mu \S^\dagger \right)
 	-\frac{f^2}{8} \str \left( \chi^\dagger \S + \S^\dagger \chi \right)
	+\frac{1}{2}m_0^2 \Phi^2
\nonumber\\&
	-a^2 C_\mathrm{Mix} \str \left( T_3 \S T_3 \S^\dagger \right)
	-W^\prime \Big[ \str \left( A^\dagger \S + \S^\dagger A \right) \Big]^2
\end{align}
where here the mass spurion includes the lattice spacing mass shift to the twisted sea mesons and is given by
\begin{align}
&\chi = \begin{pmatrix}
	\hat{m}_v \mathbf{1}_{2\times2} & & \\
	& M_s^\prime e^{i\w_0\t_3} & \\
	& & \hat{m}_v \mathbf{1}_{2\times2}
	\end{pmatrix}&
&\textrm{while}&
	&A = \begin{pmatrix} 
		0_{2\times2} && \\ & \hat{a}\, \mathbf{1}_{2\times2} & \\ &&0_{2\times2} 
	\end{pmatrix},&
\end{align}
with
\begin{align}
	&M_s^\prime = \sqrt{ (\hat{m}_{s} + \hat{a})^2 + \hat{\mu}^2},&
\nonumber\\
	&\hat{m}_v = 2B_0 m_{val},&
	&\hat{m}_s = 2B_0 m_{sea},&
	&\hat{\mu} = 2B_0 \mu,&
	&\hat{a} = 2W_0 a,&
\end{align}
The singlet field is defined as
\begin{equation}
	\Phi \equiv \frac{f}{2i} \ln \textrm{sdet} \S
	=\str \phi\, .
\end{equation}
Ultimately the singlet will be integrated out of the theory but it is convenient to keep around to determine the structure of the neutral propagators~\cite{Sharpe:2001fh}.

%
%
\subsubsection{Vacuum angle and meson masses}
We first address the shifted vacuum caused by the twisted mass term.  It is straightforward to check that the vacuum energy is minimized by expanding about
\begin{align}\label{eq:twisted_vac}
	&\S = \xi_0\, \S_{phys}\, \xi_0&
	&\textrm{with}&
	&\xi_0 = \begin{pmatrix}1_{2\times2} & &\\ & e^{i\w_0 \t_3 /2} & \\ & & 1_{2\times2} \end{pmatrix}\, .&
\end{align}
Expanding Eq.~\eqref{eq:MA_tm_Lag} around this vacuum, one then finds the valence-valence, valence-sea and sea-sea pion masses are given at arbitrary twist and LO by (we have neglected terms of $\mc{O}(m_q a)$ here which are proportional to $\cos \w$)
\begin{align}
	&(m_{\pi^{\pm,0}}^{vv})^2 = \hat{m}_{v},&
\nonumber\\
	&(m_{\pi^{\pm,0}}^{vs})^2 = \frac{1}{2}\hat{m}_{v} +\frac{1}{2}M_s^\prime +a^2 \D_\textrm{Mix}
		+\frac{32 W^\prime}{f^2}\hat{a}^2 \cos^2 \w,&
\nonumber\\
	&(m_{\pi^{\pm}}^{ss})^2 = M^\prime_s +\frac{64 W^\prime}{f^2}\hat{a}^2 \cos^2 \w,&
\nonumber\\
	&(m_{\pi^{0}}^{ss})^2 = M^\prime_s -\frac{64 W^\prime}{f^2}\hat{a}^2 \sin^2 \w
		+\frac{64 W^\prime}{f^2}\hat{a}^2 \cos^2 \w \, .&
\end{align}
We then see at maximal twist, the valence-sea mesons only receive lattice spacing corrections from the $a^2 \D_\mathrm{Mix}$ term, in agreement with Eq.~\eqref{eq:LO_phiMass}.

%
%
\subsubsection{Hairpin interactions}
We must also address the hairpin interactions.  In Ref.~\cite{Golterman:2005xa} it was shown that in addition to the Lagrangian, Eq.~\eqref{eq:MA_tm_Lag}, there are additional hairpin interactions which arise from the operators
\begin{equation}
	\d \mc{L} = -\frac{(af)^2}{32}\g_{ss} \Big[ \str \Big( \mc{P}_s(\S - \S^\dagger) \Big) \Big]^2\, .
\end{equation}
Expanding about the twisted vacuum, Eq.~\eqref{eq:twisted_vac}, this leads to an interaction
\begin{equation}\label{eq:extraHairpins}
	\d \mc{L} = \frac{1}{2} \g_{ss} a^2 \cos^2 \w\ \str \left( \mc{P}_S \phi \right)\, .
\end{equation}
Away from maximal twist, this interaction acts like a shift in the partial quenching parameters, Eq.~\eqref{eq:mod_DPQ}, although the coefficient $\g_{ss}$ would need to be determined.  However, we see that at maximal twist, this extra hairpin interaction is absent.  In the notation of Ref.~\cite{Sharpe:2000bc}, we can derive the form of the flavor neutral propagators at arbitrary twist.  First ignoring the extra hairpin, one can show the valence-valence neutral propagators are given by
\begin{equation}
\mc{G}_{v_1 v_2} = G_{v_1 v_1}^0 
	- \frac{G_{v_1 v_1}^0 V_{val} G_{v_2 v_2}}{1 + \tr \left( V_{sea} G_{ss}^0 \right) }\, ,
\end{equation}
where 
\begin{align}
	&V_{val} = \frac{m_0^2}{2} \begin{pmatrix} \ 1\ &1\ \\ \ 1\ &1\ \end{pmatrix},& 
\end{align}
and
\begin{align}
&V_{sea} = \frac{m_0^2}{2} \begin{pmatrix} \ 1\ &1\ \\ \ 1\ &1\ \end{pmatrix}
	+\frac{16W^\prime \hat{a}^2}{f^2} \begin{pmatrix}
		2 \cos^2 \w - \sin^2 \w & \sin^2 \w \\
		\sin^2 \w & 2\cos^2 \w - \sin^2 \w \end{pmatrix}\, .&
\end{align}
The valence-valence flavor neutral propagators, including the extra hairpin interactions of Eq.~\eqref{eq:extraHairpins} are then given at arbitrary twist by
\begin{equation}
\mc{G}_{v_1 v_2} = \frac{\delta_{v_1 v_2}}{p^2 + (m_\pi^{v_1 v_1})^2}
	-\frac{1}{2} \frac{p^2 + (m_\pi^{ss})^2 + \left( \frac{64W^\prime \hat{a}^2}{f^2} +2a^2 \g_{ss}\right) \cos^2 \w }
	{(p^2 + (m_\pi^{v_1 v_1})^2)(p^2 + (m_\pi^{v_2 v_2})^2)}\, .
\end{equation}
Defining the fields $\pi^0 = \frac{1}{\sqrt{2}}(\eta_u^v - \eta_d^v)$ and $\bar{\eta} = \frac{1}{\sqrt{2}}(\eta_u^v + \eta_d^v)$, one finds at maximal twist
\begin{align}
	&\mc{G}_{\pi^0} = \frac{1}{p^2 + (m_\pi^{vv})^2},& &\textrm{and}&
	&\mc{G}_{\bar{\eta}} = \frac{(m_{\pi^\pm}^{ss})^2 - (m_{\pi}^{vv})^2}
		{[\, p^2 + (m_\pi^{vv})^2\, ]^2}\, .&
\end{align}
It is interesting to note that it is not the mass of the neutral twisted mass pion which enters the numerator of the hairpin propagator, but rather a mass which is equivalent to the charged pion mass, which at maximal twist is free of even $\mc{O}(a^2)$ corrections.

%
%


\begin{thebibliography}{999}

\bibitem{Renner:2004ck}
  D.~B.~Renner {\it et al.}  [LHP Collaboration],
  Nucl.\ Phys.\ Proc.\ Suppl.\  {\bf 140} (2005) 255
  [arXiv:hep-lat/0409130].

\bibitem{Edwards:2005kw}
  R.~G.~Edwards {\it et al.}  [LHP Collaboration],
  PoS {\bf LAT2005} (2006) 056
  [arXiv:hep-lat/0509185].

\bibitem{Bowler:2004hs}
  K.~C.~Bowler, B.~Joo, R.~D.~Kenway, C.~M.~Maynard and R.~J.~Tweedie  [UKQCD
                  Collaboration],
  JHEP {\bf 0508} (2005) 003
  [arXiv:hep-lat/0411005].

\bibitem{Bonnet:2004fr}
  F.~D.~R.~Bonnet, R.~G.~Edwards, G.~T.~Fleming, R.~Lewis and D.~G.~Richards
                  [LHP Collaboration],
  Phys.\ Rev.\  D {\bf 72} (2005) 054506
  [arXiv:hep-lat/0411028].

\bibitem{Beane:2005rj}
  S.~R.~Beane, P.~F.~Bedaque, K.~Orginos and M.~J.~Savage  [NPLQCD
                  Collaboration],
  Phys.\ Rev.\  D {\bf 73} (2006) 054503
  [arXiv:hep-lat/0506013].

\bibitem{Edwards:2005ym}
  R.~G.~Edwards {\it et al.}  [LHP Collaboration],
  Phys.\ Rev.\ Lett.\  {\bf 96} (2006) 052001
  [arXiv:hep-lat/0510062].

\bibitem{Beane:2006mx}
  S.~R.~Beane, P.~F.~Bedaque, K.~Orginos and M.~J.~Savage [NPLQCD
                  Collaboration],
  Phys.\ Rev.\ Lett.\  {\bf 97} (2006) 012001
  [arXiv:hep-lat/0602010].

\bibitem{Beane:2006pt}
  S.~R.~Beane, K.~Orginos and M.~J.~Savage [NPLQCD
                  Collaboration],
  Phys.\ Lett.\  B {\bf 654} (2007) 20
  [arXiv:hep-lat/0604013].

\bibitem{Beane:2006fk}
  S.~R.~Beane, K.~Orginos and M.~J.~Savage [NPLQCD
                  Collaboration],
  Nucl.\ Phys.\  B {\bf 768} (2007) 38
  [arXiv:hep-lat/0605014].

\bibitem{Beane:2006kx}
  S.~R.~Beane, P.~F.~Bedaque, K.~Orginos and M.~J.~Savage [NPLQCD
                  Collaboration],
  Phys.\ Rev.\  D {\bf 75} (2007) 094501
  [arXiv:hep-lat/0606023].

\bibitem{Alexandrou:2006mc}
  C.~Alexandrou, T.~Leontiou, J.~W.~Negele and A.~Tsapalis,
  Phys.\ Rev.\ Lett.\  {\bf 98} (2007) 052003
  [arXiv:hep-lat/0607030].

\bibitem{Beane:2006gj}
  S.~R.~Beane, P.~F.~Bedaque, T.~C.~Luu, K.~Orginos, E.~Pallante, A.~Parreno and M.~J.~Savage [NPLQCD
                  Collaboration],
  Phys.\ Rev.\  D {\bf 74} (2006) 114503
  [arXiv:hep-lat/0607036].

\bibitem{Bar:2006zj}
  O.~Bar, K.~Jansen, S.~Schaefer, L.~Scorzato and A.~Shindler,
  PoS {\bf LAT2006} (2006) 199
  [arXiv:hep-lat/0609039].

\bibitem{Hasenfratz:2006bq}
  A.~Hasenfratz and R.~Hoffmann,
  Phys.\ Rev.\  D {\bf 74} (2006) 114509
  [arXiv:hep-lat/0609067].

\bibitem{Edwards:2006qx}
  R.~G.~Edwards {\it et al.},
  PoS {\bf LAT2006}, 121 (2006)
  [arXiv:hep-lat/0610007].

\bibitem{Beane:2006gf}
  S.~R.~Beane, P.~F.~Bedaque, T.~C.~Luu, K.~Orginos, E.~Pallante, A.~Parreno and M.~J.~Savage
                  [NPLQCD Collaboration],
  Nucl.\ Phys.\  A {\bf 794} (2007) 62
  [arXiv:hep-lat/0612026].

\bibitem{Orginos:2007tw}
  K.~Orginos and A.~Walker-Loud,
  Phys.\ Rev.\  D {\bf 77} (2008) 094505
  [arXiv:0705.0572 [hep-lat]].

\bibitem{Hagler:2007xi}
  Ph.~Hagler {\it et al.}  [LHPC Collaborations],
  Phys.\ Rev.\  D {\bf 77} (2008) 094502
  [arXiv:0705.4295 [hep-lat]].

\bibitem{Bar:2002nr}
  O.~Bar, G.~Rupak and N.~Shoresh,
  Phys.\ Rev.\  D {\bf 67}, 114505 (2003)
  [arXiv:hep-lat/0210050].

\bibitem{Bar:2003mh}
  O.~Bar, G.~Rupak and N.~Shoresh,
  Phys.\ Rev.\  D {\bf 70} (2004) 034508
  [arXiv:hep-lat/0306021].

\bibitem{Tiburzi:2005vy}
  B.~C.~Tiburzi,
  Nucl.\ Phys.\  A {\bf 761} (2005) 232
  [arXiv:hep-lat/0501020].

\bibitem{Bar:2005tu}
  O.~Bar, C.~Bernard, G.~Rupak and N.~Shoresh,
  Phys.\ Rev.\  D {\bf 72} (2005) 054502
  [arXiv:hep-lat/0503009].

\bibitem{Golterman:2005xa}
  M.~Golterman, T.~Izubuchi and Y.~Shamir,
  Phys.\ Rev.\  D {\bf 71} (2005) 114508
  [arXiv:hep-lat/0504013].

\bibitem{Tiburzi:2005is}
  B.~C.~Tiburzi,
  Phys.\ Rev.\  D {\bf 72} (2005) 094501
  [arXiv:hep-lat/0508019].

\bibitem{Chen:2005ab}
  J.~W.~Chen, D.~O'Connell, R.~S.~Van de Water and A.~Walker-Loud,
  Phys.\ Rev.\  D {\bf 73} (2006) 074510
  [arXiv:hep-lat/0510024].

\bibitem{O'Connell:2006sha}
  D.~O'Connell,
  PoS {\bf LAT2006}, 057 (2006)
  [arXiv:hep-lat/0609046].

\bibitem{Bunton:2006va}
  T.~B.~Bunton, F.~J.~Jiang and B.~C.~Tiburzi,
  Phys.\ Rev.\  D {\bf 74} (2006) 034514
  [Erratum-ibid.\  D {\bf 74} (2006) 099902]
  [arXiv:hep-lat/0607001].

\bibitem{Aubin:2006hg}
  C.~Aubin, J.~Laiho and R.~S.~Van de Water,
  Phys.\ Rev.\  D {\bf 75} (2007) 034502
  [arXiv:hep-lat/0609009].

\bibitem{Chen:2006wf}
  J.~W.~Chen, D.~O'Connell and A.~Walker-Loud,
  Phys.\ Rev.\  D {\bf 75} (2007) 054501
  [arXiv:hep-lat/0611003].

\bibitem{WalkerLoud:2006ub}
  A.~Walker-Loud,
  arXiv:nucl-th/0611048.

\bibitem{Jiang:2007sn}
  F.~J.~Jiang,
  arXiv:hep-lat/0703012.

\bibitem{Kennedy:2004ae}
  A.~D.~Kennedy,
  Nucl.\ Phys.\ Proc.\ Suppl.\  {\bf 140} (2005) 190
  [arXiv:hep-lat/0409167].

\bibitem{Edwards:2005an}
  R.~G.~Edwards, B.~Joo, A.~D.~Kennedy, K.~Orginos and U.~Wenger,
  PoS {\bf LAT2005} (2006) 146
  [arXiv:hep-lat/0510086].

\bibitem{Kaplan:1992bt}
  D.~B.~Kaplan,
  Phys.\ Lett.\  B {\bf 288} (1992) 342
  [arXiv:hep-lat/9206013].

\bibitem{Shamir:1993zy}
  Y.~Shamir,
  Nucl.\ Phys.\  B {\bf 406} (1993) 90
  [arXiv:hep-lat/9303005].

\bibitem{Furman:1994ky}
  V.~Furman and Y.~Shamir,
  Nucl.\ Phys.\  B {\bf 439} (1995) 54
  [arXiv:hep-lat/9405004].

\bibitem{Narayanan:1992wx}
  R.~Narayanan and H.~Neuberger,
  Phys.\ Lett.\  B {\bf 302} (1993) 62
  [arXiv:hep-lat/9212019].

\bibitem{Narayanan:1994gw}
  R.~Narayanan and H.~Neuberger,
  Nucl.\ Phys.\  B {\bf 443} (1995) 305
  [arXiv:hep-th/9411108].

\bibitem{Wilson:1974sk}
  K.~G.~Wilson,
  Phys.\ Rev.\  D {\bf 10} (1974) 2445.

\bibitem{Sheikholeslami:1985ij}
  B.~Sheikholeslami and R.~Wohlert,
  Nucl.\ Phys.\  B {\bf 259} (1985) 572.

\bibitem{Frezzotti:2000nk}
  R.~Frezzotti, P.~A.~Grassi, S.~Sint and P.~Weisz  [Alpha collaboration],
  JHEP {\bf 0108} (2001) 058
  [arXiv:hep-lat/0101001].

\bibitem{Kogut:1974ag}
  J.~B.~Kogut and L.~Susskind,
  Phys.\ Rev.\  D {\bf 11} (1975) 395.

\bibitem{Susskind:1976jm}
  L.~Susskind,
  Phys.\ Rev.\  D {\bf 16} (1977) 3031.

\bibitem{Ginsparg:1981bj}
  P.~H.~Ginsparg and K.~G.~Wilson,
  Phys.\ Rev.\  D {\bf 25} (1982) 2649.

\bibitem{Luscher:1998pqa}
  M.~Luscher,
  Phys.\ Lett.\  B {\bf 428} (1998) 342
  [arXiv:hep-lat/9802011].

\bibitem{Orginos:1998ue}
  K.~Orginos and D.~Toussaint  [MILC collaboration],
  Phys.\ Rev.\  D {\bf 59} (1999) 014501
  [arXiv:hep-lat/9805009].

\bibitem{Orginos:1999cr}
  K.~Orginos, D.~Toussaint and R.~L.~Sugar  [MILC Collaboration],
  Phys.\ Rev.\  D {\bf 60} (1999) 054503
  [arXiv:hep-lat/9903032].

\bibitem{Bernard:2001av}
  C.~W.~Bernard {\it et al.},
  Phys.\ Rev.\  D {\bf 64} (2001) 054506
  [arXiv:hep-lat/0104002],\\
  url: http://qcd.nersc.gov/

\bibitem{Bernard:1993sv}
  C.~W.~Bernard and M.~F.~L.~Golterman,
  Phys.\ Rev.\  D {\bf 49} (1994) 486
  [arXiv:hep-lat/9306005].

\bibitem{Sharpe:1997by}
  S.~R.~Sharpe,
  Phys.\ Rev.\  D {\bf 56} (1997) 7052
  [Erratum-ibid.\  D {\bf 62} (2000) 099901]
  [arXiv:hep-lat/9707018].

\bibitem{Sharpe:2000bc}
  S.~R.~Sharpe and N.~Shoresh,
  Phys.\ Rev.\  D {\bf 62} (2000) 094503
  [arXiv:hep-lat/0006017].

\bibitem{Sharpe:2001fh}
  S.~R.~Sharpe and N.~Shoresh,
  Phys.\ Rev.\  D {\bf 64} (2001) 114510
  [arXiv:hep-lat/0108003].

\bibitem{Weinberg:1978kz}
  S.~Weinberg,
  Physica A {\bf 96} (1979) 327.

\bibitem{Gasser:1983yg}
  J.~Gasser and H.~Leutwyler,
  Annals Phys.\  {\bf 158} (1984) 142.

\bibitem{Gasser:1984gg}
  J.~Gasser and H.~Leutwyler,
  Nucl.\ Phys.\  B {\bf 250} (1985) 465.

\bibitem{Symanzik:1983dc}
  K.~Symanzik,
  Nucl.\ Phys.\  B {\bf 226} (1983) 187.

\bibitem{Symanzik:1983gh}
  K.~Symanzik,
  Nucl.\ Phys.\  B {\bf 226} (1983) 205.

\bibitem{Sharpe:1998xm}
  S.~R.~Sharpe and R.~L.~.~Singleton,
  Phys.\ Rev.\  D {\bf 58} (1998) 074501
  [arXiv:hep-lat/9804028].

\bibitem{Labrenz:1996jy}
  J.~N.~Labrenz and S.~R.~Sharpe,
  Phys.\ Rev.\  D {\bf 54} (1996) 4595
  [arXiv:hep-lat/9605034].

\bibitem{Savage:2001jw}
  M.~J.~Savage,
  Phys.\ Rev.\  D {\bf 65} (2002) 034014
  [arXiv:hep-ph/0109190].

\bibitem{Chen:2001yi}
  J.~W.~Chen and M.~J.~Savage,
  Phys.\ Rev.\  D {\bf 65} (2002) 094001
  [arXiv:hep-lat/0111050].

\bibitem{Beane:2002vq}
  S.~R.~Beane and M.~J.~Savage,
  Nucl.\ Phys.\  A {\bf 709} (2002) 319
  [arXiv:hep-lat/0203003].

\bibitem{Kaplan:1998sz}
  D.~B.~Kaplan, M.~J.~Savage and M.~B.~Wise,
  Phys.\ Rev.\  C {\bf 59} (1999) 617
  [arXiv:nucl-th/9804032].

\bibitem{Beane:2002np}
  S.~R.~Beane and M.~J.~Savage,
  Phys.\ Rev.\  D {\bf 67} (2003) 054502
  [arXiv:hep-lat/0210046].

\bibitem{Tiburzi:2004kd}
  B.~C.~Tiburzi,
  Phys.\ Rev.\  D {\bf 71} (2005) 034501
  [arXiv:hep-lat/0410033].

\bibitem{Tiburzi:2004mv}
  B.~C.~Tiburzi,
  Phys.\ Rev.\  D {\bf 71} (2005) 054504
  [arXiv:hep-lat/0412025].

\bibitem{Mehen:2006vv}
  T.~Mehen and B.~C.~Tiburzi,
  Phys.\ Rev.\  D {\bf 74} (2006) 054505
  [arXiv:hep-lat/0607023].

\bibitem{Weinberg:1990rz}
  S.~Weinberg,
  Phys.\ Lett.\  B {\bf 251} (1990) 288.
  
\bibitem{Weinberg:1991um}
  S.~Weinberg,
  Nucl.\ Phys.\  B {\bf 363} (1991) 3.

\bibitem{Weinberg:1992yk}
  S.~Weinberg,
  Phys.\ Lett.\  B {\bf 295} (1992) 114
  [arXiv:hep-ph/9209257].

\bibitem{Kaplan:1998tg}
  D.~B.~Kaplan, M.~J.~Savage and M.~B.~Wise,
  Phys.\ Lett.\  B {\bf 424} (1998) 390
  [arXiv:nucl-th/9801034].

\bibitem{Kaplan:1998we}
  D.~B.~Kaplan, M.~J.~Savage and M.~B.~Wise,
  Nucl.\ Phys.\  B {\bf 534} (1998) 329
  [arXiv:nucl-th/9802075].

\bibitem{Beane:2002nu}
  S.~R.~Beane and M.~J.~Savage,
  Phys.\ Lett.\  B {\bf 535} (2002) 177
  [arXiv:hep-lat/0202013].

\bibitem{Rupak:2002sm}
  G.~Rupak and N.~Shoresh,
  Phys.\ Rev.\  D {\bf 66} (2002) 054503
  [arXiv:hep-lat/0201019].

\bibitem{Beane:2003xv}
  S.~R.~Beane and M.~J.~Savage,
  Phys.\ Rev.\  D {\bf 68} (2003) 114502
  [arXiv:hep-lat/0306036].

\bibitem{Munster:2003ba}
  G.~Munster and C.~Schmidt,
  Europhys.\ Lett.\  {\bf 66} (2004) 652
  [arXiv:hep-lat/0311032].

\bibitem{Scorzato:2004da}
  L.~Scorzato,
  Eur.\ Phys.\ J.\  C {\bf 37} (2004) 445
  [arXiv:hep-lat/0407023].

\bibitem{Sharpe:2004ny}
  S.~R.~Sharpe and J.~M.~S.~Wu,
  Phys.\ Rev.\  D {\bf 71} (2005) 074501
  [arXiv:hep-lat/0411021].

\bibitem{WalkerLoud:2005bt}
  A.~Walker-Loud and J.~M.~S.~Wu,
  Phys.\ Rev.\  D {\bf 72} (2005) 014506
  [arXiv:hep-lat/0504001].

\bibitem{Lee:1999zxa}
  W.~J.~Lee and S.~R.~Sharpe,
  Phys.\ Rev.\  D {\bf 60} (1999) 114503
  [arXiv:hep-lat/9905023].

\bibitem{Aubin:2003mg}
  C.~Aubin and C.~Bernard,
  Phys.\ Rev.\  D {\bf 68} (2003) 034014
  [arXiv:hep-lat/0304014].

\bibitem{Sharpe:2004is}
  S.~R.~Sharpe and R.~S.~Van de Water,
  Phys.\ Rev.\  D {\bf 71} (2005) 114505
  [arXiv:hep-lat/0409018].

\bibitem{Bailey:2007iq}
  J.~A.~Bailey,
  Phys.\ Rev.\  D {\bf 77} (2008) 054504
  [arXiv:0704.1490 [hep-lat]].

\bibitem{Aubin:2008wk}
  C.~Aubin, J.~Laiho and R.~S.~Van de Water,
  Phys.\ Rev.\  D {\bf 77} (2008) 114501
  [arXiv:0803.0129 [hep-lat]].

\bibitem{Aubin:2004fs}
  C.~Aubin {\it et al.}  [MILC Collaboration],
  Phys.\ Rev.\  D {\bf 70} (2004) 114501
  [arXiv:hep-lat/0407028].

\bibitem{Aoki:1989rx}
  S.~Aoki and A.~Gocksch,
  Phys.\ Rev.\ Lett.\  {\bf 63} (1989) 1125
  [Erratum-ibid.\  {\bf 65} (1990) 1172].

\bibitem{Guadagnoli:2002nm}
  D.~Guadagnoli, V.~Lubicz, G.~Martinelli and S.~Simula,
  JHEP {\bf 0304} (2003) 019
  [arXiv:hep-lat/0210044].

\bibitem{Berruto:2005hg}
  F.~Berruto, T.~Blum, K.~Orginos and A.~Soni,
  Phys.\ Rev.\  D {\bf 73} (2006) 054509
  [arXiv:hep-lat/0512004].

\bibitem{Shintani:2005xg}
  E.~Shintani {\it et al.},
  Phys.\ Rev.\  D {\bf 72} (2005) 014504
  [arXiv:hep-lat/0505022].

\bibitem{Shintani:2006xr}
  E.~Shintani {\it et al.},
  Phys.\ Rev.\  D {\bf 75} (2007) 034507
  [arXiv:hep-lat/0611032].

\bibitem{Aoki:1990ix}
  S.~Aoki, A.~Gocksch, A.~V.~Manohar and S.~R.~Sharpe,
  Phys.\ Rev.\ Lett.\  {\bf 65} (1990) 1092.

\bibitem{O'Connell:2005un}
  D.~O'Connell and M.~J.~Savage,
  Phys.\ Lett.\  B {\bf 633} (2006) 319
  [arXiv:hep-lat/0508009].

\bibitem{Crewther:1979pi}
  R.~J.~Crewther, P.~Di Vecchia, G.~Veneziano and E.~Witten,
  Phys.\ Lett.\  B {\bf 88} (1979) 123
  [Erratum-ibid.\  B {\bf 91} (1980) 487].

\bibitem{Jenkins:1990jv}
  E.~E.~Jenkins and A.~V.~Manohar,
  Phys.\ Lett.\  B {\bf 255} (1991) 558.

\bibitem{Jenkins:1991es}
  E.~E.~Jenkins and A.~V.~Manohar,
  Phys.\ Lett.\  B {\bf 259} (1991) 353.

\bibitem{Jenkins:1991ne}
  E.~E.~Jenkins and A.~V.~Manohar,
  \textit{Talk presented at the} Workshop on Effective Field Theories
  of the Standard Model, Dobogoko, Hungary, Aug 1991

\bibitem{Jenkins:1991ts}
  E.~E.~Jenkins,
  Nucl.\ Phys.\  B {\bf 368} (1992) 190.

\bibitem{Billeter:2004wx}
  B.~Billeter, C.~E.~Detar and J.~Osborn,
  Phys.\ Rev.\  D {\bf 70} (2004) 077502
  [arXiv:hep-lat/0406032].

\bibitem{Mueller:1998fv}
  D.~Mueller, D.~Robaschik, B.~Geyer, F.~M.~Dittes and J.~Horejsi,
  Fortsch.\ Phys.\  {\bf 42} (1994) 101
  [arXiv:hep-ph/9812448].

\bibitem{Ji:1996ek}
  X.~D.~Ji,
  Phys.\ Rev.\ Lett.\  {\bf 78} (1997) 610
  [arXiv:hep-ph/9603249].

\bibitem{Radyushkin:1997ki}
  A.~V.~Radyushkin,
  Phys.\ Rev.\  D {\bf 56} (1997) 5524
  [arXiv:hep-ph/9704207].

\bibitem{Arndt:2001ye}
  D.~Arndt and M.~J.~Savage,
  Nucl.\ Phys.\  A {\bf 697} (2002) 429
  [arXiv:nucl-th/0105045].

\bibitem{Chen:2001eg}
  J.~W.~Chen and X.~D.~Ji,
  Phys.\ Lett.\  B {\bf 523} (2001) 107
  [arXiv:hep-ph/0105197].

\bibitem{Chen:2001nb}
  J.~W.~Chen and X.~D.~Ji,
  Phys.\ Rev.\ Lett.\  {\bf 87} (2001) 152002
  [Erratum-ibid.\  {\bf 88} (2002) 249901]
  [arXiv:hep-ph/0107158].

\bibitem{Chen:2001pva}
  J.~W.~Chen and X.~D.~Ji,
  Phys.\ Rev.\ Lett.\  {\bf 88} (2002) 052003
  [arXiv:hep-ph/0111048].

\bibitem{Chen:2001gra}
  J.~W.~Chen and M.~J.~Savage,
  Nucl.\ Phys.\  A {\bf 707} (2002) 452
  [arXiv:nucl-th/0108042].

\bibitem{Beane:2004rf}
  S.~R.~Beane and M.~J.~Savage,
  Phys.\ Rev.\  D {\bf 70} (2004) 074029
  [arXiv:hep-ph/0404131].

\bibitem{Detmold:2005pt}
  W.~Detmold and C.~J.~D.~Lin,
  Phys.\ Rev.\  D {\bf 71} (2005) 054510
  [arXiv:hep-lat/0501007].

\bibitem{Huang:1957im}
  K.~Huang and C.~N.~Yang,
  Phys.\ Rev.\  {\bf 105} (1957) 767.

\bibitem{Hamber:1983vu}
  H.~W.~Hamber, E.~Marinari, G.~Parisi and C.~Rebbi,
  Nucl.\ Phys.\  B {\bf 225} (1983) 475.

\bibitem{Luscher:1986pf}
  M.~Luscher,
  Commun.\ Math.\ Phys.\  {\bf 105} (1986) 153.

\bibitem{Luscher:1990ux}
  M.~Luscher,
  Nucl.\ Phys.\  B {\bf 354} (1991) 531.

\bibitem{Beane:2003da}
  S.~R.~Beane, P.~F.~Bedaque, A.~Parreno and M.~J.~Savage,
  Phys.\ Lett.\  B {\bf 585} (2004) 106
  [arXiv:hep-lat/0312004].

\bibitem{Sato:2007ms}
  I.~Sato and P.~F.~Bedaque,
  Phys.\ Rev.\  D {\bf 76} (2007) 034502
  [arXiv:hep-lat/0702021].

\bibitem{Bedaque:2006yi}
  P.~F.~Bedaque, I.~Sato and A.~Walker-Loud,
  Phys.\ Rev.\  D {\bf 73} (2006) 074501
  [arXiv:hep-lat/0601033].

\bibitem{Beane:2001bc}
  S.~R.~Beane, P.~F.~Bedaque, M.~J.~Savage and U.~van Kolck,
  Nucl.\ Phys.\  A {\bf 700} (2002) 377
  [arXiv:nucl-th/0104030].

\bibitem{Beane:2003yx}
  S.~R.~Beane, P.~F.~Bedaque, A.~Parreno and M.~J.~Savage,
  Nucl.\ Phys.\  A {\bf 747} (2005) 55
  [arXiv:nucl-th/0311027].

\bibitem{Golterman:1997st}
  M.~F.~L.~Golterman and K.~C.~L.~Leung,
  Phys.\ Rev.\  D {\bf 57} (1998) 5703
  [arXiv:hep-lat/9711033].

\bibitem{Lin:2003tn}
  C.~J.~D.~Lin, G.~Martinelli, E.~Pallante, C.~T.~Sachrajda and G.~Villadoro,
  Phys.\ Lett.\  B {\bf 581} (2004) 207
  [arXiv:hep-lat/0308014].

\bibitem{WalkerLoud:2004hf}
  A.~Walker-Loud,
  Nucl.\ Phys.\  A {\bf 747} (2005) 476
  [arXiv:hep-lat/0405007].

\bibitem{Tiburzi:2004rh}
  B.~C.~Tiburzi and A.~Walker-Loud,
  Nucl.\ Phys.\  A {\bf 748} (2005) 513
  [arXiv:hep-lat/0407030].

\bibitem{Tiburzi:2005na}
  B.~C.~Tiburzi and A.~Walker-Loud,
  Nucl.\ Phys.\  A {\bf 764} (2006) 274
  [arXiv:hep-lat/0501018].

\bibitem{Beane:2002ca}
  S.~R.~Beane and M.~J.~Savage,
  Nucl.\ Phys.\  B {\bf 636} (2002) 291
  [arXiv:hep-lat/0203028].

\bibitem{Arndt:2003ww}
  D.~Arndt and B.~C.~Tiburzi,
  Phys.\ Rev.\  D {\bf 68} (2003) 094501
  [arXiv:hep-lat/0307003].

\bibitem{Arndt:2003we}
  D.~Arndt and B.~C.~Tiburzi,
  Phys.\ Rev.\  D {\bf 68} (2003) 114503
  [Erratum-ibid.\  D {\bf 69} (2004) 059904]
  [arXiv:hep-lat/0308001].

\bibitem{Arndt:2003vd}
  D.~Arndt and B.~C.~Tiburzi,
  Phys.\ Rev.\  D {\bf 69} (2004) 014501
  [arXiv:hep-lat/0309013].

\bibitem{Detmold:2006vu}
  W.~Detmold, B.~C.~Tiburzi and A.~Walker-Loud,
  Phys.\ Rev.\  D {\bf 73} (2006) 114505
  [arXiv:hep-lat/0603026].

\bibitem{Chen:2006gg}
  J.~W.~Chen, W.~Detmold and B.~Smigielski,
  Phys.\ Rev.\  D {\bf 75} (2007) 074003
  [arXiv:hep-lat/0612027].

\bibitem{Detmold:2006gh}
  W.~Detmold and C.~J.~D.~Lin,
  Phys.\ Rev.\  D {\bf 76} (2007) 014501
  [arXiv:hep-lat/0612028].

\bibitem{Sharpe:2004ps}
  S.~R.~Sharpe and J.~M.~S.~Wu,
  Phys.\ Rev.\  D {\bf 70}, 094029 (2004)
  [arXiv:hep-lat/0407025].

\end{thebibliography}
\end{document}